\documentclass[journal, 10pt]{IEEEtran}
\IEEEoverridecommandlockouts
\usepackage{cite}
\usepackage{amsmath,amssymb,amsfonts}
\usepackage{algorithm}
\usepackage[noend]{algorithmic}
\usepackage{graphicx}
\usepackage{textcomp}
\usepackage{xcolor}
\usepackage{xspace}
\usepackage{cleveref}
\usepackage{multirow}
\usepackage{multicol}
\usepackage{url}
\usepackage{booktabs}
\usepackage{enumitem}

\newtheorem{theorem}{Theorem}

\def\BibTeX{{\rm B\kern-.05em{\sc i\kern-.025em b}\kern-.08em
    T\kern-.1667em\lower.7ex\hbox{E}\kern-.125emX}}
\begin{document}

\newcommand{\ours}{AVR\textsubscript{dp}\xspace}
\newcommand{\avr}{AVR\xspace}
\newcommand{\pono}{Pono\xspace}
\newcommand{\ia}{IC3IA\xspace}
\newcommand\todo[1]{ {\textcolor{red}{TODO: #1}} }
\newcommand\zap[1]{}

\title{Leveraging Datapath Propagation in IC3 for Hardware Model Checking}

\author{
\IEEEauthorblockN{Hongyu Fan\IEEEauthorrefmark{1}, Fei He\IEEEauthorrefmark{2}\\ 
\thanks{Fei He is corresponding author}}
\IEEEauthorblockA{
 School of Software, Tsinghua University, Key Laboratory for Information System Security, MoE \\
 Beijing National Research Center for Information Science and Technology, Beijing, China \\
 Email: fhy18@mails.tsinghua.edu.cn \ \ hefei@mail.tsinghua.edu.cn 
}
}

\maketitle

\begin{abstract}
IC3 is a famous bit-level framework for safety verification. 
By incorporating datapath abstraction, a notable enhancement 
in the efficiency of hardware verification can be achieved. 
However, datapath abstraction entails a coarse level of 
abstraction where all datapath operations are approximated 
as uninterpreted functions. This level of abstraction, 
albeit useful, can lead to an increased computational burden 
during the verification process as it necessitates extensive 
exploration of redundant abstract state space.

In this paper, we introduce a novel approach called datapath 
propagation. Our method involves leveraging concrete constant 
values to iteratively compute the outcomes of relevant datapath 
operations and their associated uninterpreted functions. 
Meanwhile, we generate potentially useful datapath propagation 
lemmas in abstract state space and tighten the datapath abstraction. 
With this technique, the abstract state space can be reduced, 
and the verification efficiency is significantly improved. 
We implemented the proposed approach and conducted extensive 
experiments. The results show promising improvements of our 
approach compared to the state-of-the-art verifiers. 
\end{abstract}

\begin{IEEEkeywords}
datapath abstraction, datapath propagation, hardware verification, reachability safety
\end{IEEEkeywords}

\section{Introduction}
\label{sec:introduction}

IC3 algorithm~\cite{ic3-11} (also known as PDR \cite{pdr-11}) is arguably
the most talented and successful technique for verifying reachability 
safety property in hardware model checking. The IC3 algorithm incrementally 
explores the design's state space and tries to construct a proof of correctness. 
However, the bit-level IC3 algorithms \cite{abc-10, avy-12} suffer from 
the state space explosion problem. As the bit-width and complexity 
of the hardware design increase, the IC3 algorithm's performance 
degenerates rapidly, threatening its scalability. 

In Verilog RTL design, the datapath is responsible for processing and 
manipulating data as it flows through the system. Datapath operations 
refer to the specific procedures performed within the datapath, 
typically involves arithmetic, logical, and data movement operations. 
They are the most essential units in the design. 
Complex functions usually consist of many datapath operations.
A popular approach~\cite{avr-14, avr-19} integrates the IC3 algorithm 
with datapath abstraction~\cite{dp-abstract95, dp-abstract-thesis}, 
which denotes datapath operations as uninterpreted functions (UF) 
instead of their explicit implementation details. Therefore, 
UFs serve as over-approximations for datapath operations. 
Importantly, the verification process does not entail defining 
the specific logic or functionality encapsulated within these functions. 
This abstraction facilitates a higher level of generality, enabling 
a more abstract representation of the overall datapath behavior.

Constraint solving for UF \cite{smt-02, SMTLIB2} is much faster than 
bit-vector (BV) \cite{barrett1998decision, barrett2010smt,sovlingBV07}. 
Therefore, each IC3 call on abstract state space is mostly 
several orders of magnitude more efficient than the bit-level IC3 call. 
However, abstraction may bring spurious counterexamples. For each 
abstract counterexample returned by IC3, the \emph{counterexample-guided 
abstraction refinement} (CEGAR)~\cite{cegar-00, cegar-03} procedure 
checks if the abstract counterexample is spurious. If it is, then 
the refinement procedure generates \emph{datapath refinement lemmas} 
to prune the abstract state space and tighten the current 
abstraction. Each datapath lemma is a constraint formula over UFs. 
Then, the verification procedure calls the IC3 iteratively 
until either the property holds or a real counterexample is found. 

Although integrating the IC3 with datapath abstraction and refinement 
is shown to be successful and practical \cite{avr-tacas20, I4-SOSP}, 
the iteration of the CEGAR is also a crucial factor to the overall 
efficiency of the verification process. However, the CEGAR is designed 
as a general framework for abstraction-based verification. 
The knowledge of datapath operations cannot be fully utilized. 
Roughly abstracting all the datapath operations as uninterpreted 
functions makes the verification procedure lose all the semantics 
of datapath operations. Constraint solving can assign arbitrary 
values to UFs. Therefore, the verification procedure may find numerous 
spurious counterexamples, even if they are trivial, and require 
many rounds of refinement to tighten the abstraction. On the other 
hand, applying this knowledge in CEGAR may be useful for pruning 
the abstract state space and thus improving the overall efficiency. 

Our basic idea is to utilize the knowledge of datapath operations in 
abstract state space by propagating the accurate values to corresponding 
UFs and guide the datapath abstraction-based IC3 for hardware verification. 
A straightforward attempt for this idea is constant propagation 
\cite{constant86, wegman1991constant, cp2017}, an optimization technique 
that aims to identify and propagate constant values throughout a system. 
It replaces variables or expressions with their known constant values, 
eliminating unnecessary computations and improving runtime performance. 
However, constant propagation is performed in the concrete state space. 

In this paper, we propose a \emph{datapath propagation} that 
propagates constant values from concrete state space to abstract 
state space. In more detail, for each abstract constraint formula, 
we first recognize constant values and related UFs. 
Then, we consider the original semantics of UFs, i.e., their 
corresponding datapath operations. We propagate constant values 
to drive the outcomes of these datapath operations and assign 
the outcomes to the corresponding UFs. Subsequently, we substitute 
these UFs with their outcomes and continue the iterative propagation. 
With this technique, some UFs are assigned with accurate values, 
or relations between UFs and constants can be determined. 
Therefore, we prune redundant abstract space and tighten the 
datapath abstraction.

Moreover, we propose to generate \emph{datapath propagation lemma} 
(DPL), which is another type of datapath lemma generated during 
the propagation process. It can be generated in two situations. 
First, once the propagation deduces that the current formula 
is unsatisfiable, we generate DPL to record the core reason 
and block the possible spurious counterexample. Second, if some 
predicates or binary relations referring to datapath operations 
are determined during the propagation, we generate DPL to record 
this information and facilitate further verification. 
The generated DPLs can also eliminate spurious counterexamples 
and, more importantly, reduce the number of CEGAR iterations. 
Our method is performed in the abstract state space, independent 
of CEGAR. Combining our approach and refinement in CEGAR can further 
prune the abstract state space. The verification efficiency is thus improved. 

We implemented the proposed method on top of \avr~\cite{avr-tacas20}, 
which is the champion tool of the latest Hardware Model Checking 
Competition (HWMCC). Our implementation is called \ours. 
We conducted experiments on 1089 benchmarks collected from HWMCC
2019 and 2020 -- the last two competitions. We compare \ours with 
state-of-the-art hardware verification tools, including \avr, 
\ia~\cite{ic3ia-14}, and \pono~\cite{pono21} with four different 
engines. The experimental results show that \ours solves 87, 479, 
545-727 more cases than \avr, \ia, and \pono, respectively. 
Counting on both-verified cases, \ours achieves 1.46x, 20.04x, 
1.13x-11.46x speedup over \avr, \ia, and \pono, respectively. 
Especially, \ours generates 3923 datapath propagation lemmas 
and reduces 29.5\% refinements than \avr. 

The contributions of this paper are summarized as follows:
\begin{itemize}
    \item We proposed a novel datapath propagation approach 
          in abstraction-based IC3 for hardware verification. 
    \item We devised a datapath propagation lemma generation 
          procedure, with which the deduced results can be 
          kept to facilitate further verification. 
    \item We implemented the proposed method on top of \avr 
          and conducted extensive experiments to evaluate 
          its effectiveness and efficiency. Experimental 
          results show the promising performance of our approach.
\end{itemize}

The rest of this paper is organized as follows. Section~\ref{sec:preliminary} 
introduces necessary preliminaries. Section~\ref{sec:motivation} uses 
a example to motivate our approach. Section\ref{sec:propagation} details 
the datapath propagation. Experimental results and analysis are presented 
in Section~\ref{sec:evaluation}, followed by related works in 
Section~\ref{sec:related} and conclusion in Section~\ref{sec:conclusion}.
\section{Preliminaries}
\label{sec:preliminary}

\subsection{Notations}

In \emph{first-order logic} (FOL), a \emph{term} is a variable, a constant, 
or an $n$-ary function applied to $n$ terms; an \emph{atom} is $\bot$, 
$\top$, or an $n$-ary predicate applied to $n$ terms; 
a \emph{literal} is an \emph{atom} or its negation. 
A \emph{cube} is a conjunction of literals and a \emph{clause} 
is a disjunction of literals. A \emph{first-order formula} 
is built from literals using Boolean connectives and quantifiers.
An interpretation (or model) $M$ consists of a non-empty object 
set $\mathit{dom}(M)$, called the \emph{domain} of $M$,
an assignment that maps each variable to an object in $\mathit{dom}(M)$,
and an interpretation for each constant, function, and predicate, respectively. 
A formula $\Phi$ is \emph{satisfiable} if there exists a model $M$ so that 
$M \models \Phi$; $\Phi$ is \emph{valid} if for any model $M$, $M \models \Phi$.

\begin{figure}[tb]
  \centering
  \includegraphics[width=.9\linewidth]{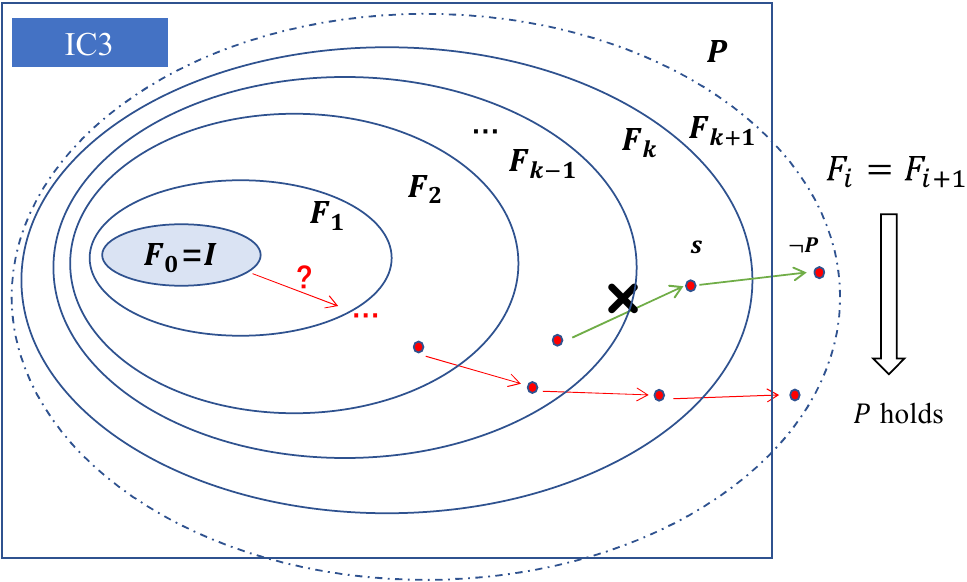}
  \caption{Overview of the IC3 algorithm.}
  \label{fig:ic3}
\end{figure}

A \emph{first-order theory} $\mathcal{T}$ is defined by a \emph{signature} 
and a set of \emph{axioms}. The \emph{signature} consists of constant symbols, 
function symbols, and predicate symbols allowed in $\mathcal{T}$; 
the \emph{axioms} prescribe the intended meanings of these symbols.
A \emph{$\mathcal{T}$-model} is a model that satisfies all axioms 
of $\mathcal{T}$. A formula $\Phi$ is \emph{$\mathcal{T}$-satisfiable} 
if there exists a $\mathcal{T}$-model $M$ so that $M \models \Phi$; 
$\Phi$ is \emph{$\mathcal{T}$-valid} if it is satisfied by all 
$\mathcal{T}$-models. The \emph{satisfiability modulo theories} 
(SMT) \cite{smt-02,smt-11,SMTLIB2} problem is a decision problem 
for formula $\phi$ in some combination of first-order theories. 
For each theory $\mathcal{T}$ in $\phi$, there is a $\mathcal{T}$-solver 
that can check the $\mathcal{T}$-satisfiability of conjunctions 
of literals in $\mathcal{T}$. 

\subsection{Model Checking}

A hardware design can be represented as a netlist or a model 
in a hardware description language such as Verilog. 
Let $X$ be the set of state variables in the design; 
let $X'$ be the primed copy of $X$ representing the next-state variables. 
The design's behavior can be encoded as a model checking 
problem via a 4-tuple $\mathcal{P} := \langle X, I, T, P\rangle$, where 
$I(X)$ is a formula for the initial states, 
$T(X, X')$ is a formula for the transition relation,  
and $P(X)$ is a formula for the desired safety property. 
Specifically, the next-state variables in $T$ are represented 
as functions of present-state variables. Input variables 
are treated as state variables whose next states are unconstrained. 

A \emph{state} $s$ is an assignment to all variables in $X$.
A $trace$ is a sequence of states $s_0, s_1, \dots, s_k$ such that 
$I(s_0)$ holds, and $T(s_i, s_{i+1})$ holds for $0 \leq i \leq k-1$. 
The property formula $P$ asserts that all reachable states 
satisfy $P$, i.e., $P$ should be invariant for the design. 
Otherwise, there must be a finite trace $s_0, s_1, \dots, s_k$, 
which is a counterexample that $P(s_k)$ does not hold. 
An \emph{inductive invariant} $F$ is a formula satisfying: 
(1) $I \Rightarrow F$, and (2) $F$ is closed under 
the transition relation, i.e., $F \wedge T \Rightarrow F'$. 

\begin{algorithm}[htb]
  \begin{algorithmic}[1]
    \IF{$I \wedge \neg P$ or $I \wedge T \wedge \neg P'$}
      \STATE{return counterexample trace;}
    \ENDIF
    \STATE{$k = 1, F_k = P$;}
    \WHILE{true}
    {
      \WHILE{$F_k \wedge T \wedge \neg P'$}
      {
        \STATE{let $s$ be the satisfying assignment;}
        \IF{Reachable$(s, I)$}
          \STATE{return counterexample trace;}
        \ELSE
          \STATE{Block$(s, k+1)$;}
        \ENDIF
      }
      \ENDWHILE
      \IF{$F_i = F_{i-1}$ for some $2 \leq i \leq k+1$}
        \STATE{return empty trace; // $P$ hold}
      \ENDIF
      \STATE{$k$++;}
    }
    \ENDWHILE
  \end{algorithmic}
  \caption{IC3 ($I,T,P$)}
  \label{alg:ic3}
\end{algorithm}

\subsection{IC3 Algorithm}

IC3 (or PDR) is a well-known algorithm for determining whether 
a hardware design satisfies a given safety property $P$. 
It represents a major advance over previous SAT-based 
induction methods \cite{Bjesse20,Sheeran20,McMillan03}. 
Fig.~\ref{fig:ic3} shows an overview of the IC3 algorithm. 
It maintains a sequence of frontiers $F_0, F_1, \dots, 
F_k$ where $F_0 = I$ and $F_i, i>0$ is an over-approximation 
of reachable states after $i$ steps from $I$. Suppose a state 
$s$ in $F_k$ violates $P$ after a 1-step transition; the 
algorithm tries constructing a trace that witnesses the violation. 
A counterexample is returned if $s$ is reachable from $I$ (red line). 
Otherwise, $s$ is eliminated (green line) by tightening $F_0, F_1, 
\dots, F_k$. Once two frontier approximations equal, IC3 returns 
an empty trace proving that $P$ is satisfied. 

Alg.~\ref{alg:ic3} lists the pseudo-code of IC3. Taking $I, T, P$ as input, 
it first looks for 0-step and 1-step counterexample traces (line 1). 
If none are found, the algorithm instantiates $F_k$, the over-approximation 
of $k$-step reachable states ($k \geq 1$), to $P$ (line 3). Then, 
the IC3 algorithm iteratively checks if $F_k$ can reach $\neg P$ 
states in one transition (line 6). Each satisfying assignment $s$ 
is checked to determine if it is reachable from $I$ (line 7). 
If unreachable, e.g., the green trace in Fig.~\ref{fig:ic3}, 
$s$ is blocked and used to tighten the frontiers $F_1$ to $F_{k+1}$ 
(line 10). This check continues until either a counterexample trace 
of the length $k+1$ is found (line 8), or all the states that violate 
$P$ in one transition are unreachable from $I$. If the algorithm 
finds $F_i = F_{i-1}$ for some $2 \leq i \leq k+1$, it returns an 
empty trace indicating that $P$ holds (line 11), and $F_i$ is an 
inductive invariant that satisfies $P$ ($F_k \Rightarrow P$). 
Otherwise, the IC3 algorithm increments $k$ and continues to check 
the existence of counterexample traces on longer transitions (line 13). 

The pseudo-code shows a sketch of the algorithm and hides many details. 
It consists of numerous 1-step backward reachability checks processed 
in order. These reachability checks are represented as formulas using 
the BV theory. Therefore, the state space is exponential in the bit 
width of state variables. Given $\mathcal{P}$ with $n$ total bits, 
each bit can be 0 or 1, and there are up to $2^n$ states in the concrete 
state space. Therefore, as the bit width increases, the scale of the state 
space grows exponentially, and the efficiency of IC3 degenerates rapidly. 

\subsection{Datapath Abstraction}

Abstraction is a common technique for improving the efficiency and scalability 
of verification. It creates an abstract model that captures the critical behavior 
and properties of the system while approximating certain details as needed. 
The abstract model brings higher-level representation and simplified views 
of the original system. In this way, an abstract state can represent a cluster 
of concrete states. Therefore, it reduces the proof of a property on an 
infinite or large concrete state space to a proof on an abstract state space. 

Datapath abstraction \cite{avr-14} replaces state variables and 
datapath operations with uninterpreted functions (UF). It returns the abstract 
version of the original problem $\mathcal{P} := \langle X, I, T, P\rangle$ 
as $\hat{\mathcal{P}} := \langle \hat{X}, \hat{I}, \hat{T}, \hat{P}\rangle$. 
$\hat{\mathcal{P}}$ over-approximates the original system and is a sound 
abstraction, i.e., if $\hat{P}$ is proved safe on the abstract state space, 
so is $P$ on the concrete state space. However, a counterexample that 
violates $\hat{P}$ on the abstract state space may be spurious on the 
concrete state space due to the coarse abstraction. 
\section{Motivation}
\label{sec:motivation}

\begin{figure}[t]
  \centering
  \includegraphics[width=.75\linewidth]{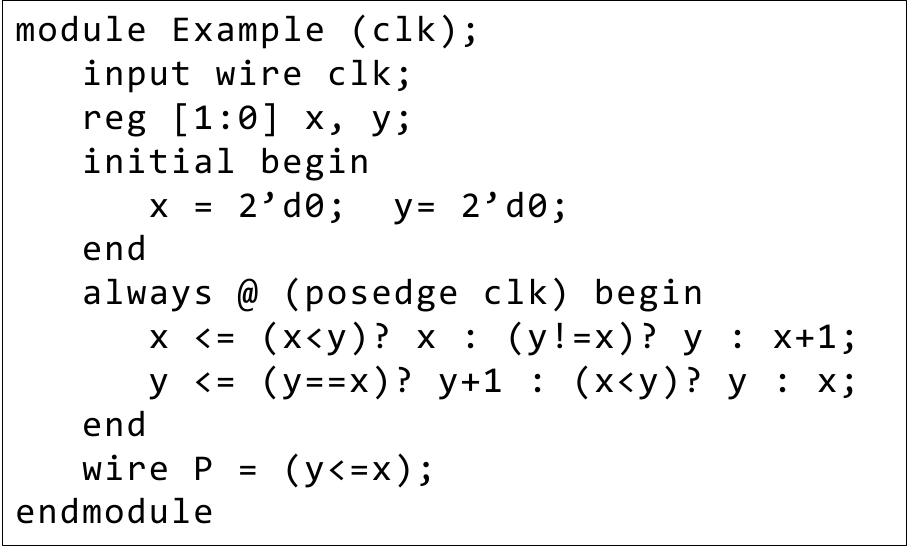}
  \caption{Verilog description of an example sequential circuit with 
  a specified safety property. The state variables are 2-bit unsigned 
  integers $x=x_1x_0$ and $y=y_1y_0$. The main sequential logic involves 
  computing the next-state values of $x$ and $y$. The safety property 
  asserts that $y \leq x$ is always satisfied.}
  \label{fig:example}
\end{figure}

In this section, we use a simple example to motivate our approach. 
We first introduce the integration of datapath abstraction and IC3. 
Then, we show that some important knowledge of datapath operations 
is neglected, and utilizing this knowledge can be quite useful. 

\subsection{IC3 with Datapath Abstraction}

Consider the example design in Fig.~\ref{fig:example}. The design's 
behavior can be encoded as a model checking problem $\mathcal{P}$ 
where $x, y$ are state variables and $I, T, P$ are:
\begin{equation*}
  \begin{split}
    I : \ &x = 0 \ \wedge \ y = 0 \\
    T : \ &x' = (x<y) \ ?\ x : (y \neq x) \ ?\ y :x+1 \ \wedge \ \\ 
        \  &y' = (y==x)\ ?\ y+1 : (x < y)\ ?\ y:x \\
    P : \ &y \leq x
  \end{split}
\end{equation*}

IC3 can be enhanced with the datapath abstraction. The enhanced algorithm, 
called DP-IC3\cite{avr-14}, is shown in Alg.~\ref{alg:dp-ic3}. It first 
calls DP-Abstract to perform the datapath abstraction (line 1). 
For the example design, we have: 
\begin{equation*}
  \begin{split}
    \hat{I} : \ &\hat{x}=\hat{0} \ \wedge \ \hat{y}=\hat{0} \\
    \hat{T} : \ &\hat{x}' = LT(\hat{x},\hat{y}) \ ?\ \hat{x} : (\hat{y} \neq 
                            \hat{x})\ ?\ \hat{y} : ADD(\hat{x},\hat{1}) \ \wedge \ \\ 
              \ &\hat{y}' = (\hat{y}==\hat{x}) \ ?\ ADD(\hat{y},\hat{1}) 
                            : LT(\hat{x}, \hat{y}) \ ?\ \hat{y} : \hat{x} \\
    \hat{P} : \ & LE(\hat{y},\hat{x})
  \end{split}
\end{equation*}
$\hat{\mathcal{P}}$ uses uninterpreted sort and converts datapath 
operations (e.g., $<, +$) with UFs (e.g., LT, ADD). Note that state 
variables or constants are denoted as 0-ary UFs. 

Then, IC3 runs on the abstract state space. Let $\Phi_{drl}$ be the 
conjunction of datapath refinement lemmas derived in CEGAR. It is 
initialized to $true$ (line 2). Lines 3-12 are the main body of DP-IC3. 
It calls the modified version of IC3 in Alg.~\ref{alg:ic3} that operates 
on abstract formulas. Note that $\Phi_{drl}$ serves as the fourth 
argument and augments all the queries that IC3 performs. If IC3 returns 
an empty trace, no counterexample is found in the abstract or concrete 
state space. Alg.~\ref{alg:dp-ic3} terminates with the conclusion that 
$P$ holds (lines 5-6). 

Otherwise, a non-empty trace representing an abstract counterexample 
(ACEX) is found. DP-IC3 calls DP-Concrete to generate CEX as the bit-level 
version of ACEX and checks its feasibility (line 8). If CEX is feasible, 
Alg.~\ref{alg:dp-ic3} returns CEX as a counterexample trace that 
witnesses the violations of $P$ (line 10). If CEX is infeasible, 
Alg.~\ref{alg:dp-ic3} calls DP-Refine to eliminate the spurious 
counterexample by generating datapath refinement lemmas (line 13). 
Then, DP-IC3 invokes the next round of IC3. 

Consider the example design in Fig.\ref{fig:example}, IC3 first checks 
0-step safety by calling an SMT solver. $\hat{I}$ is safe iff the query 
formula $\hat{I} \wedge \neg \hat{P}$ is unsatisfiable. 
A 0-step abstract counterexample $acex_1$ is returned: 
\[acex_1: \{\hat{x} \mapsto \hat{0}, \ \hat{y} \mapsto \hat{0}, \ 
  \textit{LE}(\hat{y}, \hat{x}) \mapsto \textit{false} \}\]
where $a \mapsto b$ means $b$ is the assignment of $a$ by the SMT solver. 
To check its feasibility, DP-concrete returns $cex_1$ as its bit-level 
counterpart: $x = 0 \wedge y = 0 \wedge y > x$. Then, $cex_1$ is 
bit-blasted and feasibility checking is performed using the BV theory 
in SMT solver. Apparently, $cex_1$ is BV-unsatisfiable. DP-IC3 realizes 
that $acex_1$ is spurious and derives the datapath refinement lemma 
$drl_1: \textit{LE}(\hat{0}, \hat{x})$ by calling DP-Refine. This lemma 
is added to $\Phi_{drl}$, which tightens the abstraction and prevents 
$acex_1$ from appearing again. 

In the second iteration, IC3 checks 1-step safety ($\hat{I} \wedge \hat{T} 
\neg \hat{P}' \wedge drl_1$) and returns a 1-step ACEX $acex_2$. We don't 
show the trace for brevity. The corresponding bit-level formula $cex_2$:
\[x=0 \wedge y=0 \wedge x \geq y \wedge x' = x+1 \wedge y' = y+1 \wedge y'>x'\]
is found to be infeasible by the SMT solver. Then, the refinement procedure 
refutes $acex_2$ by generating datapath refinement lemma $drl_2$:
\begin{equation*}
  \begin{split}
    \neg ( \hat{x}=\hat{0} \ \wedge \ & \hat{y}=\hat{0} \wedge
           \hat{x}'=\textit{ADD}(\hat{x},\hat{1}) \ \wedge \\ 
          &\hat{y}'=\textit{ADD}(\hat{y},\hat{1}) \wedge 
           \neg \textit{LE}(\hat{y}',\hat{x}'))
  \end{split}
\end{equation*} 

\begin{algorithm}[htb]
  \begin{algorithmic}[1]
    \STATE{$\hat{I}, \hat{T}, \hat{P}$ = DP-Abstract($I,T,P$);}
    \STATE{$\Phi_{drl}  =true$; // initialize datapath lemmas}
    \WHILE{true}
    {
      \STATE{ACEX = IC3($\hat{I}, \hat{T}, \hat{P}, \Phi_{drl}$);}
      \IF{ACEX is empty}
        \STATE{return empty trace; // $P$ holds}
      \ELSE
        \STATE{CEX = DP-Concrete(ACEX);}
        \IF{CEX is feasible}
        \STATE{return CEX; // $P$ fails}
        \ENDIF
      \ENDIF
      \STATE{$\Phi_{drl} = \Phi_{drl} \ \wedge \ $DP-Refine(ACEX);}
    }
    \ENDWHILE
  \end{algorithmic}
  \caption{DP-IC3 ($I,T,P$)}
  \label{alg:dp-ic3}
\end{algorithm}

The third iteration also returns a 1-step ACEX $acex_3$, which is 
found to be infeasible and refuted by datapath refinement lemma 
$drl_3: \neg (\hat{y} = \hat{x} \wedge \textit{LT}(\hat{x}, \hat{y}))$.  
Finally, after six refinements, DP-IC3 succeeds in finding an 
inductive invariant $\hat{y}=\hat{x}$ and proves that $\hat{P}$ 
(and $P$) holds. 

\subsection{Datapath Knowledge is Important}

The inherent advantage of the DP-IC3 is that the reachability 
computation is performed on the abstract model of the hardware 
design, which hides the bit-level details of datapath operations. 
From the angle of the SMT solver, constraint solving for UFs is 
much faster than when bit-level facts must be involved. Therefore, 
each IC3 call is expected to be more efficient than the bit-level 
IC3 call. However, the bit-level IC3 is only called once, but 
DP-IC3 may call IC3 iteratively in abstract state space because 
of the spurious counterexamples. Therefore, the number of CEGAR 
iterations is curial to the overall verification efficiency. 

Applying the knowledge of datapath operations can reduce the 
CEGAR iterations. Consider the 0-step safety check of the example 
design in Fig.~\ref{fig:example}. Since predicate $\textit{LE}$ 
is uninterpreted for SMT solver, and there are no other constraints 
on it, $\textit{LE}(\hat{y}, \hat{x})$ can be assigned to any 
boolean values, which results in the spurious counterexample 
$acex_1$. Each CEGAR iteration is complex. DP-IC3 needs to generate 
the bit level counterexample $cex_1$, checking its feasibility, 
then invokes DP-Refine to generate datapath refinement lemmas 
that refute $acex_1$. In contrast, $0 \leq 0$ (even $0 \leq x$) 
is trivial for $\leq$, considering its semantics. Therefore, 
convey the information that $\textit{LE}(\hat{0}, \hat{0})$ 
or $\textit{LE}(\hat{0}, \hat{x})$ to DP-IC3 is useful for 
avoiding the spuriousness and reducing CEGAR iterations. 

Moreover, applying the knowledge datapath operations can reduce 
the size of the query formula. In DP-IC3, every query formula is 
augmented by $\Phi_{drl}$. As the size of $\Phi_{drl}$ grows 
along with the CEGAR iterations, the size of the query formula 
also grows rapidly. Therefore, reducing the CEGAR iterations 
can also reduce the size of the query formula in abstract space 
and achieves higher constraint solving efficiency. 

However, the knowledge of datapath operations is neglected by DP-IC3. 
Datapath operations are the essential arithmetic or logical units 
in the design which comprise complex functionalities. Roughly treating 
all datapath operations as UFs causes coarse abstraction, which may 
bring numerous spurious counterexamples and put a heavy burden on the 
CEGAR framework. Instead, for a query formula $\hat{\varphi}$ in abstract 
state space, utilizing the knowledge of datapath operations may prune 
the redundant search space, reduce the number of CEGAR iterations, 
and improve the verification efficiency. 

One may consider all the semantics of datapath operations. Then, DP-IC3 
degenerates to the bit-level IC3 algorithm since no abstraction exists  
and it suffers from the state space exploration problem. To take 
advantage of DP-IC3 and utilize the knowledge of datapath operations, 
we propose a datapath propagation procedure. 
\section{Datapath Propagation}
\label{sec:propagation}

This section details the datapath propagation. We first introduce the supported 
datapath operations and an informal description of the propagation. Then, we show 
the propagation rules that carry the knowledge of datapath operations. Finally, 
we show the workflow of the propagation procedure. 

\subsection{Overview}

Let $\varphi$ be a bit-level formula in \emph{conjunction normal form} (CNF), 
e.g., $\varphi = C_1 \wedge C_2 \wedge \dots C_n$ where $C_i, 0 \leq i \leq n$, 
is a clause. Denote $\alpha$ and $\gamma$ as the DP-Abstract and DP-Concrete 
functions in Alg.~\ref{alg:dp-ic3}. Therefore, $\alpha$ replaces constants, 
variables, or datapath operations with UFs; $\gamma$ is just the opposite. 
$\alpha$ and $\gamma$ maintain the correspondence between an abstract entity 
and its bit-level counterpart in datapath abstraction. Let $u$ be a constant, 
a variable, or a CNF formula, we use $\hat{u} = \alpha(u)$ to represent 
its abstract version and $u = \gamma(\hat{u})$. Tab.~\ref{tab:dp-operations} 
lists the abstract version of supported datapath operations, which can be 
divided into several categories: 
\begin{itemize}
    \item \textbf{Arithmetic Operations}: mathematical computations on data, 
          including addition (+), subtraction (-), multiplication ($\times$), 
          division ($/$), and modulo ($\%$) operations.
    \item \textbf{Relational Operations}: data comparison operations that 
          returns a Boolean value, including $<$ and $\leq$. We use their 
          negation to represent $\geq$ and $>$ to reduce the types and 
          facilitate the further analysis.  
    \item \textbf{Bitwise Operations}: logical operations manipulate binary 
          data using Boolean logic, including Bitwise AND ($\&$), Bitwise OR 
          ($|$), Bitwise XOR ($^\wedge$), Bitwise NOT ($\sim$), and negations 
          of the first three operations. 
    \item \textbf{Reduction Operations}: logical operations that reduce a set 
          of data elements to a Boolean value based on a specific operation or 
          function, including reduction AND, OR, XOR, and their negation.
    \item \textbf{Shift Operations}: data movement operations that shift the 
          binary representation of data to the left or right, including: 
          logical left shift ($<<$) and logical right shift ($>>$) where 
          the empty bits are filled with zero; arithmetic left shift ($<<<$) 
          and arithmetic right shift ($>>>$) where the sign bit is used to 
          fill the empty bit positions. 
\end{itemize}

Let $symb(\hat{\varphi})$ be a set of symbols appearing in $\hat{\varphi}$. 
For example, suppose that $\hat{\varphi}$ is $\hat{x}=\hat{0} \wedge 
\hat{y} = \textit{ADD}(\hat{x}, \hat{1}) \wedge LT(\hat{y}, \hat{x})$
we have $symb(\hat{\varphi}) = \{\hat{x}, \hat{y}, \hat{0}, \hat{1}, 
\textit{LT}, \textit{ADD}\}$. 

The main idea of datapath propagation is to first recognize constant symbols in 
$\hat{\varphi}$, and propagates them to related UFs. We consider the original 
datapath operations of these UFs and try to obtain the outcomes of the original 
datapath operations. Moreover, the knowledge of datapath operations may beyond 
the constant symbols, we also devise the propagation rules to utilize this 
knowledge and obtain the outcomes of related datapath operations. Then we assign 
the outcomes back to the corresponding UFs and continue the iterative propagation. 
In the following, we introduce the propagation rules. 

\begin{table}
    \centering
    \caption{Abstract datapath operations}
    \begin{tabular}{cl}
    \toprule
    \multicolumn{1}{c}{Type}       & \multicolumn{1}{c}{After DP-Abstrcat}  \\ \cmidrule(lr){1-2}
    \multicolumn{1}{c}{Arithmetic} & ADD, SUB/Minus, MUL, DIV, MOD          \\ \cmidrule(lr){1-2}
    \multicolumn{1}{c}{Relational} & LT, LE                                 \\ \cmidrule(lr){1-2}
    \multicolumn{1}{c}{Bit-wise}   & 
        \begin{tabular}[c]{@{}l@{}}
            BitWiseAnd, BitWiseOr, BitWiseXor, BitWiseNAnd, \\ 
            BitWiseNor, BitWiseXNor, BitWiseNot
        \end{tabular} \\ \cmidrule(lr){1-2}
    \multicolumn{1}{c}{Reduction}  & 
        \begin{tabular}[c]{@{}l@{}}
            ReductionAnd, ReductionOr, ReductionXor,\\ 
            ReductionNAnd, ReductionNor, ReductionXNor
        \end{tabular}  \\ \cmidrule(lr){1-2}
    \multicolumn{1}{c}{Shift}      & ShiftL, ShiftR, AShiftL, AShiftR       \\
    \bottomrule
    \end{tabular}
    \label{tab:dp-operations}
\end{table}

\subsection{Propagation Rules}

Propagation rules vary with the datapath operation's type. Let $\hat{x},\hat{y},\hat{z}$ 
be 0-ary UFs that represent abstract state variables after datapath abstraction. 
Denote 0-ary UFs $\hat{0}, \hat{1},\dots$ the constant symbols. Let $\textit{MAX}_x$ 
be $2^{x.width()}-1$. Tab.~\ref{table:propagation-rules} lists the essential 
propagation rules for supported datapath operations. 

The first row shows the propagation rules for arithmetic operations. We take 
$\textit{MUL}(\hat{x}, \hat{y}) = \alpha(x \times y)$ as an example:
\begin{itemize}
    \item if $\hat{x}$ is not equal to any constant symbol in $symb(\hat{\varphi})$ 
          and $\hat{y}=\hat{0}$, $\textit{MUL}(\hat{x}, \hat{y})$ is propagated 
          to $\hat{0}$.
    \item if $\hat{x}$ is not equal to any constant symbol in $symb(\hat{\varphi})$ 
          and $\hat{y}=\hat{1}$, $\textit{MUL}(\hat{x}, \hat{y})$ is propagated to 
          that $\hat{x}$.
\end{itemize}
The other two symmetric cases have the same result. 
Moreover, $\hat{\varphi}$ is a CNF formula, i.e., $\hat{\varphi} = \hat{C_1}\wedge 
\hat{C_2} \wedge \dots \hat{C_n}$. If $\hat{\varphi}$ is satisfiable, both $\hat{C_i}, 
0 < i\leq n$ should be satisfiable. Therefore, for some $\hat{C_i}$ that are 
equalities between UFs, we put these UFs into a set (called \emph{equality closure}). 
\subsubsection*{\textup{\textbf{Example1}}}
For example, suppose $\hat{\varphi}$ is 
\[\hat{x}=\hat{y} \wedge \hat{u}=\textit{SUB}(\hat{x}, \hat{y}) \wedge \hat{v}=\hat{0} 
\wedge \textit{LT}(\hat{u},\hat{v}) \wedge \dots\]
we maintain equality closures \{$\hat{x}, \hat{y}$\} and \{$\hat{v}, \hat{0}$\} 
to facilitate further propagations. Consider the propagation rules about arithmetic 
operations in Tab.~\ref{tab:dp-operations}, $\hat{x}$ and $\hat{y}$ belong to the 
same equality closure, then $\textit{SUB}(\hat{x}, \hat{y})$ should be $\hat{0}$. 
Since $\hat{0} \in symb(\hat{\varphi})$, we have $\hat{u} = \hat{0}$ and we add 
$\hat{u}$ to the second equality closure. Since $\hat{u} = \hat{v}$, the propagation 
rule about relational operations in Tab.~\ref{tab:dp-operations} can be applied. 
We have $\neg \textit{LT}(\hat{u}, \hat{v})$, which contradicts the predicates 
$\textit{LT}(\hat{u}, \hat{v})$ in $\hat{\varphi}$. 

Note that the parameters of a UF can be more than just abstract variables or constant 
symbols, e.g., $\hat{x}=\hat{y} \rightarrow \textit{LE}(\textit{ADD}(\hat{x},\hat{1}), 
\textit{ADD}(\hat{y}, \hat{1}))$. Secondly, there are some special rules for reduction 
operations. For example, suppose $\hat{x}$ is determined to be unequal to $\hat{0}$ 
in some $\hat{C_i}$ and $\textit{ReductionOr}(\hat{x})$ appears in $\hat{\varphi}$, 
we replace $\textit{ReductionOr}(\hat{x})$ with $\hat{1}$ if $\hat{1} \in symb(\hat{\varphi})$. 
Therefore, the propagation builds equality between the UF and the constant symbol, 
which is unknown to the SMT solver. 

\begin{table*}[htb]
    \centering
    \caption{Propagation rules for arithmetic, relational, bit-wise, reduction, and shift operations}
    \begin{tabular}{ll}
    \toprule
    \multicolumn{1}{c}{Type}       & \multicolumn{1}{c}{Propagation rules} \\ 
    \cmidrule(lr){1-2}
    
    Arithmetic & 
    \begin{tabular}[l]{@{}l@{}}
        \ \ \ \ $\textit{ADD}(\hat{x}, \hat{0}) = \hat{x}$ \ \ \ 
        \ \ \ \ $\textit{ADD}(\hat{0}, \hat{y}) = \hat{y}$ \ \ \ 
        \ \ \ \ $\textit{SUB}(\hat{x}, \hat{0}) = \hat{x}$ \ \ \ 
        \ \ \ \ $\hat{x}=\hat{y} \rightarrow \textit{SUB}(\hat{x}, \hat{y}) = \hat{0}$ \ \ \ 
        \ \ \ \ $\textit{MUL}(\hat{x}, \hat{0}) = \hat{0}$ 
        \\ \addlinespace  
        \ \ \ \ $\textit{MUL}(\hat{0}, \hat{y}) = \hat{0}$ \ \ \ 
        \ \ \ \ $\textit{MUL}(\hat{x}, \hat{1}) = \hat{x}$ \ \ \ 
        \ \ \ \ $\textit{MUL}(\hat{1}, \hat{y}) = \hat{y}$ \ \ \
        \ \ \ \ $\hat{x}=\hat{y} \rightarrow \textit{DIV}(\hat{x},\hat{y})=\hat{1}$  \ \ \ 
        \ \ \ \ $\textit{DIV}(\hat{0}, \hat{y}) = \hat{0}$ 
        \\ \addlinespace
        \ \ \ \ $\textit{MOD}(\hat{x}, \hat{1}) = \hat{0}$ \ \ \  
        \ \ \ \ $\textit{MOD}(\hat{0}, \hat{x}) = \hat{0}$ \ \ \ 
        \ \ \ \ $\hat{x} = \hat{y} \rightarrow \textit{MOD}(\hat{x}, \hat{y}) = \hat{0}$
    \end{tabular}  \\ 
    \cmidrule(lr){1-2}
    
    Relational & 
    \begin{tabular}[l]{@{}l@{}}
        \ \ \ \ $\neg \textit{LT}(\hat{x}, \hat{x})$ \ \
        \ \ \ \ $\hat{x} = \hat{y} \rightarrow \neg \textit{LT}(\hat{x}, \hat{y})$ \ \ \ \ \ \ \ \
        \ \ \ \ $\textit{LT}(\hat{x}, \hat{z}) \wedge \textit{LT}(\hat{z}, \hat{y}) \rightarrow \textit{LT}(\hat{x}, \hat{y})$ \ \ 
        \ \ \ \ $\textit{LT}(\hat{x}, \hat{y}) \rightarrow \textit{LE}(\hat{x}, \hat{y})$ 
        \\ \addlinespace

        \ \ \ \ $\neg \textit{LT}(\hat{x}, \hat{0})$ \ \ 
        \ \ \ \ $\textit{LE}(\hat{x}, \hat{y}) \rightarrow \neg \textit{LT}(\hat{y}, \hat{x})$ \ \ \ \ \ 
        \ \ \ \ $\textit{LE}(\hat{x}, \hat{z}) \wedge \textit{LE}(\hat{z}, \hat{y}) \rightarrow \textit{LE}(\hat{x}, \hat{y})$ \ \
        \ \ \ \ $\textit{LT}(\hat{x}, \hat{z}) \wedge \textit{LE}(\hat{z}, \hat{y}) \rightarrow \textit{LT}(\hat{x}, \hat{y})$ \ \ \  
        \\ \addlinespace

        \ \ \ \ $\textit{LE}(\hat{0}, \hat{x})$ \ \ \ \ 
        \ \ \ \ $\hat{x} = \hat{y} \rightarrow \textit{LE}(\hat{x}, \hat{y})$ \ \ \ \ \ \ \ \ \
        \ \ \ \ $\textit{LE}(\hat{x}, \hat{z}) \wedge \textit{LE}(\hat{z}, \hat{y}) \rightarrow \textit{LE}(\hat{x}, \hat{y})$ 
    
        \\ \addlinespace
        \ \ \ \ $\textit{LE}(\hat{x}, \hat{x})$ \ \ \ 
        \ \ \ \ $\textit{LT}(\hat{x}, \hat{y}) \rightarrow \neg \textit{LE}(\hat{y}, \hat{x})$ \ \ \ \ \
        \ \ \ \ $\textit{LE}(\hat{x}, \hat{z}) \wedge \textit{LT}(\hat{z}, \hat{y}) \rightarrow \textit{LT}(\hat{x}, \hat{y})$ 
        
    \end{tabular} \\ 
    \cmidrule(lr){1-2}
    
    Bit-wise & 
    \begin{tabular}[l]{@{}l@{}}
        \ \ \ \ $\textit{BitWiseAnd}(\hat{x}, \hat{0}) = \hat{0}$ \ \ \ \ \  
        \ \ \ \ $\textit{BitWiseOr}(\hat{x}, \hat{0}) = \hat{x}$ \ \ \ \ \ \ \ \ \ \ 
        \ \ \ \ $\textit{BitWiseNor}(\hat{\textit{MAX}_x}, \hat{y}) = \hat{0}$ \ \ \
        \ \ \ \ $\hat{x}=\hat{y} \rightarrow \textit{BitWiseAnd}(\hat{x},\hat{y}) = \hat{x}$
        \\ \addlinespace

        \ \ \ \ $\textit{BitWiseAnd}(\hat{0}, \hat{y}) = \hat{0}$ \ \ \ \ \ 
        \ \ \ \ $\textit{BitWiseOr}(\hat{0}, \hat{y}) = \hat{y}$ \ \ \ \ \ \ \ \ \ \ \ 
        \ \ \   $\textit{BitWiseOr}(\hat{x}, \hat{\textit{MAX}_x}) = \hat{\textit{MAX}_x}$
        \ \ $\hat{x}=\hat{y} \rightarrow \textit{BitWiseOr}(\hat{x},\hat{y}) = \hat{x}$ \ \  
        \\ \addlinespace

        \ \ \ \ $\textit{BitWiseAnd}(\hat{x}, \hat{\textit{MAX}_x}) = \hat{x}$ 
        \ \ \ $\textit{BitWiseOr}(\hat{\textit{MAX}_x}, \hat{y}) = \hat{\textit{MAX}_x}$
        \     $\textit{BitWiseNAnd}(\hat{x}, \hat{0}) = \hat{\textit{MAX}_x}$  \
        \ \ \ \ $\hat{x}=\hat{y} \rightarrow \textit{BitWiseXor}(\hat{x},\hat{y}) = \hat{0}$
        \\ \addlinespace

        \ \ \ \ $\textit{BitWiseAnd}(\hat{\textit{MAX}_x}, \hat{y}) = \hat{y}$  
        \ \ \ $\textit{BitWiseNor}(\hat{x}, \hat{\textit{MAX}_x}) = \hat{0}$  \ \
        \ \ \ \ $\textit{BitWiseNAnd}(\hat{0}, \hat{y}) = \hat{\textit{MAX}_x}$ \
        \ \ \ \ $\hat{x}=\hat{y} \rightarrow \textit{BitWiseXNor}(\hat{x},\hat{y}) = \hat{\textit{MAX}_x}$
    \end{tabular} \\ 
    \cmidrule(lr){1-2}

    Reduction & 
    \begin{tabular}[l]{@{}l@{}}
        \ \ \ \ $\textit{ReductionAnd}(\hat{x})=\hat{0}$ where $\neg (\hat{x}=\hat{\textit{MAX}_x})$ \ \
        \ \ \ \ $\textit{ReductionOr}(\hat{x})=\hat{1}$ where $\neg (\hat{x}=\hat{0})$ \ \ \
        \ \ \ \ $\textit{ReductionNOr}(\hat{x})=\hat{1}$ where $\neg (\hat{x}=\hat{0})$
        \\ \addlinespace 
        \ \ \ \ $\textit{ReductionNAnd}(\hat{x})=\hat{0}$ where $\neg (\hat{x}=\hat{\textit{MAX}_x})$ 
        \ \ \ \ $\textit{ReductionXNOr}(\hat{x})=\hat{1}$ where $\neg (\hat{x}=\hat{\textit{MAX}_x)}$ \\
    \end{tabular} \\ 
    \cmidrule(lr){1-2}

    Shift & 
    \begin{tabular}[l]{@{}l@{}}
        \ \ \ \ $\textit{ShiftL}(\hat{x},\hat{0})=\hat{x}$ \ \ \ 
        \ \ \ \ $\textit{ShiftL}(\hat{0},\hat{x})=\hat{0}$ \ \ \ \
        \ \ \ \ $\textit{ShiftR}(\hat{x},\hat{0})=\hat{x}$ \ \ \ \
        \ \ \ \ $\textit{ShiftR}(\hat{0},\hat{x})=\hat{0}$ \\ \addlinespace
        \ \ \ \ $\textit{AShiftL}(\hat{x},\hat{0})=\hat{x}$ \ 
        \ \ \ \ $\textit{AShiftL}(\hat{0},\hat{x})=\hat{0}$ \ \ 
        \ \ \ \ $\textit{AShiftR}(\hat{x},\hat{0})=\hat{x}$ \ \ 
        \ \ \ \ $\textit{AShiftR}(\hat{0},\hat{x})=\hat{0}$ \\
    \end{tabular} \\ 
    
    \bottomrule
    \end{tabular}
    \label{table:propagation-rules}
\end{table*}

\begin{figure}[htb]
    \centering
    \includegraphics[width=\linewidth]{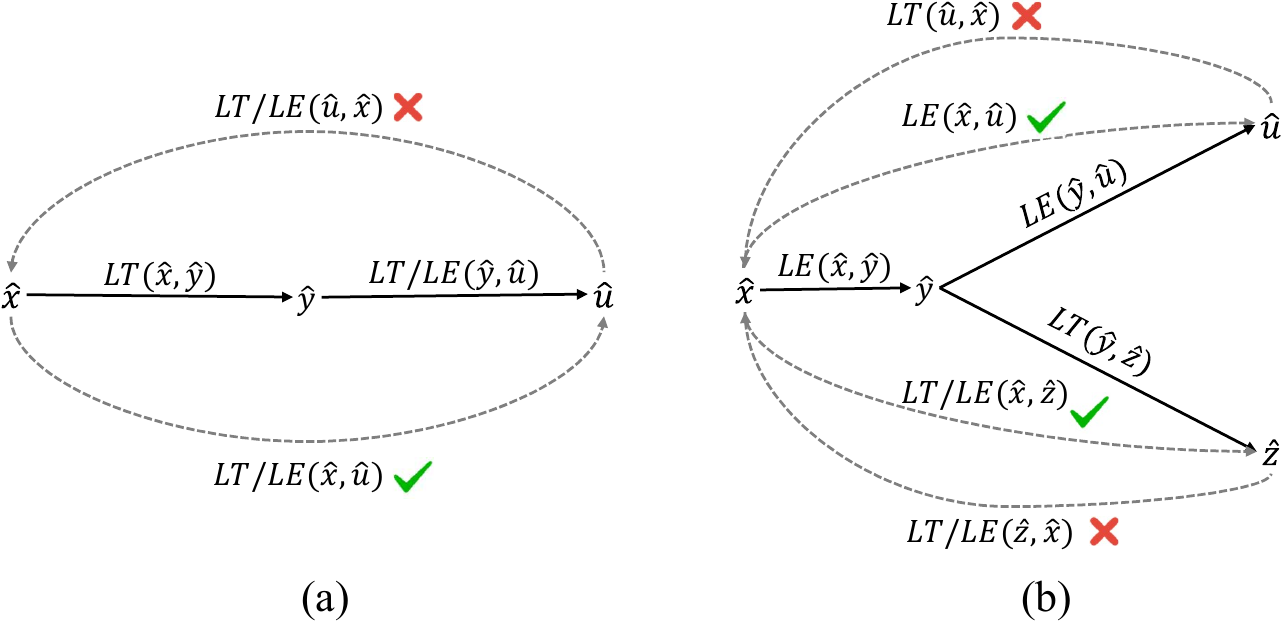}
    \caption{Propagation rules about relational operations}
    \label{fig:relational-graph}
\end{figure}

Thirdly, there are some special rules for relational operations. Consider the second 
row in Tab.~\ref{tab:dp-operations}, if $\textit{LT}$ and $\textit{LE}$ appear in 
$\hat{\varphi}$ and they have the same set of parameters, we have propagation rules 
$\textit{LT}(\hat{x},\hat{y}) \rightarrow \neg \textit{LE}(\hat{y}, \hat{x})$ since 
$x < y$ means $\neg(y \leq x)$ and $\textit{LE}(\hat{x},\hat{y}) \rightarrow \neg 
\textit{LT}(\hat{y}, \hat{x})$ since $x \leq y$ means $\neg(y < x)$. Moreover, $<$ 
and $\leq$ are transitive; Fig.~\ref{fig:relational-graph} provides an intuitive 
illustration of these propagation rules. In $(\mathtt{a})$, suppose that some 
$\hat{C_i}$ in $\hat{\varphi}$ are predicates $\textit{LT}(\hat{x}, \hat{y})$ 
and $\textit{LT}/\textit{LE}(\hat{y}, \hat{u})$ (the solid line), and 
$\textit{LT}/\textit{LE}(\hat{x}, \hat{u})$ appears in $\hat{\varphi}$. 
Since $x<y \wedge y < u$ implies $x<u$ and $x<y \wedge y \leq u$ implies $x<u$, 
we propagate $\textit{LT}(\hat{x},\hat{u})$ and $\textit{LE}(\hat{x}, \hat{u})$ 
to $\textit{true}$ (the green checkmark). On the contrary, the predicates 
$\textit{LT}(\hat{u},\hat{x})$ and $\textit{LE}(\hat{u},\hat{x})$ are propagated 
to $\textit{false}$ (the red cross) if they appear in $\hat{\varphi}$. 

In contrast, consider $\textit{LE}(\hat{x}, \hat{y})$ and $\textit{LE}(\hat{y},\hat{u})$ 
in $(\mathtt{b})$. Since $x \leq y \wedge y \leq u$ implies $x\leq u$, we propagate 
$\textit{LE}(\hat{x},\hat{u})$ to $\textit{true}$ (the green checkmark) and 
$\textit{LT}(\hat{u},\hat{x})$ just the opposite (the red cross). Note that 
$\textit{LT}(\hat{x},\hat{u})$ is still unknown if it appears in $\hat{\varphi}$. 
Moreover, since $x \leq y \wedge y < z$ implies $x < z$, the bottom half of 
$(\mathtt{b})$ has the same propagation results as $(\mathtt{a})$. 

\subsection{Propagation Procedure}

Given an abstract CNF formula $\hat{\varphi}$, \emph{datapath propagation} is a procedure 
that tries to construct a formula $\hat{\psi}$, s.t., $\hat{\varphi} \wedge \hat{\psi}$ 
is \emph{unsat} and $\models_{\mathcal{T}} \psi$ where $symb(\hat{\psi}) \subseteq 
symb(\hat{\varphi})$ and $\psi=\gamma{\hat{(\psi)}}$. In more detail, $\hat{\psi}$ 
is a formula over UFs and $\psi =\gamma(\hat{\psi})$ its bit-level counterpart. 
$\models_{\mathcal{T}} \psi$ means $\psi$ is a tautology under first order theory 
$\mathcal{T}$, e.g., bit-vector theory. Datapath propagation tries to find a formula 
$\hat{\psi}$ over UFs that $\hat{\varphi} \wedge \hat{\psi}$ is unsatisfiable. 
$symb(\hat{\psi})\subseteq symb(\hat{\varphi})$ indicates that all the symbols 
appeared in $\hat{\psi}$ also appear in $\hat{\varphi}$, i.e., $\hat{\psi}$ does 
not introduce new uninterpreted constants, variables, predicates, or functions 
that are not in $symb(\hat{\varphi})$. 

\subsubsection*{\textup{\textbf{Example2}}}
Let $\hat{\varphi}$ be $\hat{x} = \hat{0} \wedge \hat{y} = \textit{ADD}(\hat{x}, \hat{1}) 
\wedge LT(\hat{y}, \hat{x})$, a possible assignment returned by the SMT solver is:
\[\{\hat{x} \mapsto \hat{0}, \ \hat{y} \mapsto v_1, \ \textit{ADD}(\hat{x}, 
\hat{1}) \mapsto v_1 \ \textit{LT}(\hat{y}, \hat{x}) \mapsto \textit{true}\}\]
where $v_1$ can be any uncertainty value that does not exceed the maximum 
value in its bit width. In contrast, datapath propagation may find a formula 
$\hat{\psi}: \textit{ADD}(\hat{0}, \hat{1}) = \hat{1} \wedge \neg \textit{LT}(\hat{1}, 
\hat{0})$. It is easy to see that $\hat{\varphi} \wedge \hat{\psi}$ is \emph{unsat} 
and the bit-level formula $\psi: 0+1 = 1 \wedge \neg (1<0)$ is tautology under 
BV theory. Meanwhile, $symb(\hat{\psi}) \subseteq symb(\hat{\varphi})$. 

For each query formula, datapath propagation terminates in three situations: 
1) success in finding formula $\hat{\psi}$; 2) no more propagation can be conducted; 
3) failure to find $\hat{\psi}$ within the specific number of iterations. 
For the first situation, since $\psi$ is a bit-level fact and $\hat{\varphi} 
\wedge \hat{\psi}$ is unsatisfiable, $\hat{\psi}$ is a datapath lemma that 
eliminates spurious ACEX. For the last two situations, no more propagation 
can be applied, or the size of $\hat{\varphi}$ is too big to finish the 
propagation, then $\hat{\varphi}$ is passed to 
the SMT solver. However, some bit-level facts $\lambda$ can also be propagated 
in the last two situations. $\hat{\lambda} = \alpha(\lambda)$ is a datapath lemma that 
reduces the abstract state space and facilitates further verification. 
We call $\hat{\psi}$ or $\hat{\lambda}$ \emph{datapath propagation lemma} (DPL). 

Alg.~\ref{alg:dp-propagation} details the datapath propagation. The inputs include  
an abstract query formula $\hat{\varphi}$, a maximal propagation depth $\textit{bound}$, 
and a formula $\Phi_{dpl}$ that records datapath propagation lemmas (DPLs). 
The algorithm tries to construct a formula $\hat{\psi}$ that $\models_{\mathcal{T}} \psi$ 
and $\hat{\varphi}\wedge \hat{\psi}$ is \emph{unsat}. It returns $\textit{unsat}$ if successful, 
or $\textit{unknown}$ in other situations. First, we set two formulas $\hat{\varphi}_p 
= \hat{\varphi}$ and $\hat{\varphi}_q = \textit{true}$. They are used to check if 
propagation continues. $k=0$ is the current depth of the propagation, and $\hat{\psi}$ 
is initially assigned $\textit{true}$ (line 1). 

The while loop in lines 2-24 is the main procedure, which exits when $\hat{\varphi}_p 
= \hat{\varphi}_q$ or $k \geq bound$ (line 2). Since $\hat{\varphi}_q$ is the query 
formula before executing the loop and $\hat{\varphi}_p$ is the query formula after 
the loop, $\hat{\varphi}_p = \hat{\varphi}_q$ means no propagation is conducted 
in the last iteration. $k \geq bound$ means the number of iterations exceeds the 
specific depth. In the loop, we first assigns $\hat{\varphi}_p$ to $\hat{\varphi}_q$ 
(line 3) and let $\hat{C}$ be the set of constant symbols in $\hat{\varphi}$ (line 4). 

For each constant symbol $\hat{c} \in \hat{C}$, we propagate $\hat{c}$ to related 
UFs in $\hat{\varphi}$ and use $\textit{UF}_c$ to collect the updated UFs (line 6). 
In more detail, suppose that $\hat{\varphi}$ is 
\[\hat{x}=\hat{0} \wedge \hat{y}=\hat{x} \wedge \hat{z}=\hat{1} \wedge 
  \textit{ADD}(\hat{u}, \hat{y}) \wedge \textit{LT}(\hat{x}, \hat{z})\] 
propagating $\hat{0}$ to related UFs in $\hat{\varphi}$ returns $\hat{z}=\hat{1} 
\wedge \textit{ADD}(\hat{u}, \hat{0}) \wedge \textit{LT}(\hat{0}, \hat{z})$. 
$\textit{ADD}(\hat{u}, \hat{0})$ and $\textit{LT}(\hat{0}, \hat{z})$ 
are added to $\textit{UF}_c$. 

For each $\textit{uf} \in \textit{UF}_c$ (line 7), if all the parameters of 
$\textit{uf}$ are constant symbols; the algorithm considers the original datapath 
operation $\gamma(\textit{uf})$ and tries to obtain its outcome. Let $\textit{res}$ 
represents the outcome and $\hat{\textit{res}} = \alpha(\textit{res})$ its abstract 
version (line 9). If $\alpha(\textit{res})$ exists in $symb(\hat{\varphi})$, 
we update $\hat{\varphi}$ by replacing $\textit{uf}$ with $\textit{res}$ (line 11). 
Moreover, we build an equality between $\hat{\textit{res}}$ and $\textit{uf}$ 
and combines the equality into $\hat{\psi}$ (line 12). However, if $\hat{\textit{res}}$ 
introduces new symbol that dose not exists in $symb(\hat{\varphi})$, we combine 
inequalities $\neg (\textit{uf} = \hat{u})$ into $\hat{\psi}$ for each $\hat{u}$ 
in $C$ (line 14). For example, suppose $\textit{ADD}(\hat{x},\hat{y})$ appears 
in $\hat{\varphi}$ and we know that $\hat{x}=\hat{1}$ and $\hat{y}=\hat{1}$. 
Since $1+1=2$, we check if $\hat{2} \in symb(\hat{\varphi})$. If so, we replace 
$\textit{ADD}(\hat{x},\hat{y})$ with $\hat{2}$ in $\hat{\varphi}$ and combine 
$\textit{ADD}(\hat{1},\hat{1})=\hat{2}$ into $\hat{\psi}$. If only $\hat{0}, 
\hat{1}$, and $\hat{3}$ appear in $\hat{\varphi}$, then we combine 
$\neg(\textit{ADD}(\hat{1},\hat{1})=\hat{0})$, $\neg (\textit{ADD}(\hat{1},
\hat{1})=\hat{1})$, and $\neg (\textit{ADD}(\hat{1},\hat{1})=\hat{3})$ 
into $\hat{\psi}$. 

If at least one parameter is not a constant symbol in $\textit{uf}$ (line 15), we may 
not obtain the value outcome of $\gamma(\textit{uf})$. In this situation, we apply 
propagation rules described in the last section, which consider the original semantics 
of supported datapath operations (line 16). If the updated formula $\hat{\varphi} \wedge 
\hat{\psi}$ is \emph{unsat}, $\hat{\psi}$ is enough to demonstrate that $\varphi$ is 
$\hat{\mathcal{T}}$-unsatisfiable. Then, the algorithm combines $\hat{\psi}$ to $\Phi_{dpl}$ 
and returns $\textit{unsat}$ (lines 17-19). Otherwise, the algorithm continues to pick 
the next $\textit{uf} \in \textit{UF}_c$ (line 7). If all the UFs in $\textit{UF}_c$ 
are processed, the algorithm continues to operate the next constant symbol in $\hat{C}$ 
(line 5). In summary, the nested loop from lines 5 to 19 propagates constant symbols 
to related UFs. 

\begin{algorithm}
    \begin{algorithmic}[1]
        \STATE{$\hat{\varphi}_p = \hat{\varphi}, \ 
            \hat{\varphi}_q = \textit{true}, \ k = 0$, \ $\hat{\psi} = \textit{true}$;}
        \WHILE{$\hat{\varphi}_p \ != \ \hat{\varphi}_q$ and $k < \textit{bound}$}
        {
            \STATE{$\hat{\varphi}_q = \hat{\varphi}_p$;}
            \STATE{let $\hat{C}$ be constant symbols in $\hat{\varphi}_p$;}
            \FOR{$\hat{c} \in \hat{C}$}
                \STATE{$\textit{UF}_c$ = update\_related\_UF($\hat{c}, \hat{\varphi}_p$);}
                \FOR{$\textit{uf} \in \textit{UF}_c$}
                    \IF{parameters are all constant symbols in $\textit{uf}$}
                        \STATE{$\textit{res} = \gamma(\textit{uf}), \ 
                                \hat{\textit{res}} = \alpha(\textit{res})$;}
                        \IF{$\hat{\textit{res}} \in symb(\hat{\varphi}_p)$}
                            \STATE{$\hat{\varphi}_p$ = replace\_UF\_with\_const($\hat{\varphi}_p, 
                                                        \textit{uf}, \hat{\textit{res}}$);}
                            \STATE{$\hat{\psi} = \hat{\psi} \wedge (\textit{uf} = \hat{\textit{res}})$;}
                        \ELSE
                            \STATE{$\hat{\psi} = \hat{\psi} \wedge \neg (\textit{uf}  = \hat{u})$ 
                                    \textbf{foreach} $\hat{u} \in \hat{C}$;} 
                        \ENDIF
                    \ELSE
                        \STATE{$\hat{\psi}, \hat{\varphi}_p$ = apply\_propagation\_rule($\textit{uf}$);}
                    \ENDIF
                    \IF{$\hat{\varphi}_p \wedge \hat{\psi}$ is \emph{unsat}}
                        \STATE{$\Phi_{dpl} = \Phi_{dpl} \wedge \hat{\psi}$;}
                        \RETURN{$\textit{unsat}$;}
                    \ENDIF
                \ENDFOR
            \ENDFOR
            \STATE{$\hat{\psi}, \hat{\varphi}_p$ = apply\_propagation\_rule($\hat{\varphi}_p$);}
            \IF{$\hat{\varphi}_p \wedge \hat{\psi}$ is \emph{unsat}}
                \STATE{$\Phi_{dpl} = \Phi_{dpl} \wedge \hat{\psi}$;}
                \RETURN{$\textit{unsat}$;}
            \ENDIF
            \STATE{$k = k +1$;}
        }
        \ENDWHILE
        \STATE{$\Phi_{dpl} = \Phi_{dpl} \wedge \hat{\psi}$;}
        \RETURN{$\textit{unknown}$;}
    \end{algorithmic}
\caption{Datapath propagation ($\hat{\varphi}, \textit{bound}, \Phi_{dpl}$)}
\label{alg:dp-propagation}
\end{algorithm}

Next, the algorithm performs more propagations beyond constant symbols using propagation rules 
in Tab.~\ref{table:propagation-rules} (line 20). If $\hat{\varphi} \wedge \hat{\psi}$ is 
\emph{unsat}, the algorithm combines $\hat{\psi}$ to $\Phi_{dpl}$ and returns \emph{unsat} 
(lines 21-23); Otherwise, $k = k + 1$ and it continues the next round of datapath propagation 
(line 24). Finally, if $\hat{\varphi}_p = \hat{\varphi}_q$, i.e., no more propagations 
can be conducted, or $k$ exceeds the maximal depth, Alg.~\ref{alg:dp-propagation} combines 
$\hat{\psi}$ to $\Phi_{dpl}$ and returns $\textit{unknown}$ (lines 25-26). Then, $\hat{\varphi}$ 
and $\Phi_{dpl}$ are passed to the SMT solver. 

Datapath propagation iteratively utilizes the above propagation rules and tries to obtain 
the outcomes of related UFs. We have the following theorem:
\vspace{0.5\baselineskip}
\begin{theorem}
    \emph{Datapath propagation is sound and incomplete.} 
\end{theorem}
\vspace{0.5\baselineskip}
The theorem is straightforward. Propagation rules are originated from the original semantics 
of datapath operations. Datapath propagation applies these rules to UFs, which is apparently 
a sound procedure. For completeness, since we only support part of datapath operations and 
iterations are not exhaustive, datapath propagation is a incomplete procedure. 
\section{Applying Datapath Propagation in DP-IC3} 

\begin{figure}[t]
    \centering
    \includegraphics[width=\linewidth]{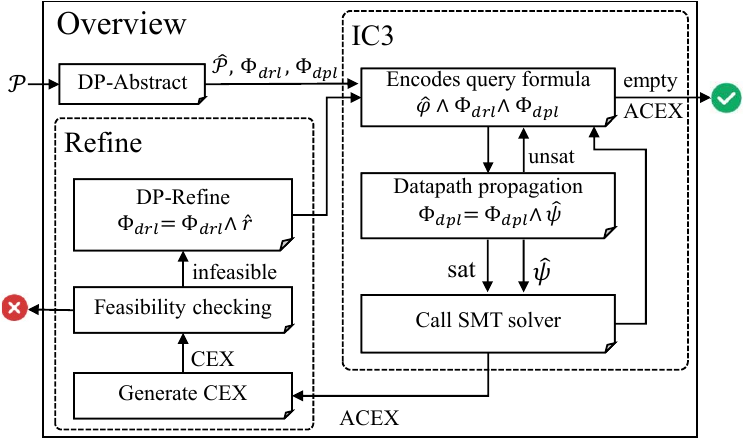}
    \caption{Overview of the DP-IC3 with datapath propagation}
    \label{fig:overview}
\end{figure}

Fig.~\ref{fig:overview} shows a high-level overview of the DP-IC3 with datapath propagation. 
Taking a model checking problem $\mathcal{P} := \langle X, I, T, P\rangle$ as input, 
we obtain $\hat{\mathcal{P}}$ after DP-Abstract and initialize $\Phi_{\textit{dpl}}$ 
and $\Phi_{\textit{drl}}$ to $\textit{true}$, which record the datapath propagation 
lemmas and datapath refinement lemmas, respectively. Then, IC3 is performed in the 
abstract space. For each query $\hat{\varphi}$, it encodes the query formula 
$\hat{\varphi} \wedge \Phi_{\textit{dpl}} \wedge \Phi_{\textit{drl}}$ and invokes 
datapath propagation. If the propagation returns $\textit{unsat}$, IC3 continues to 
encode the next abstract query formula as needed. Otherwise, IC3 combines newly 
generated DPL $\hat{\psi}$ to the query formula and passes it to the SMT solver. 
Then, IC3 continues until either it returns an empty trace indicating that $\hat{P}$ 
holds (the green checkmark) or it returns a non-empty trace ACEX. 

ACEX is passed to the refinement procedure. It first generates the bit-level 
counterpart CEX and performs feasibility checking. If CEX is feasible, a real 
counterexample is found, and $P$ is violated (the red cross). Otherwise, CEX 
is BV-unsatisfiable, DP-Refine generates datapath refinement lemmas $\hat{r}$ 
and combines $\hat{r}$ to $\Phi_{\textit{drl}}$ to eliminate ACEX. 
Then, the verification framework calls the next round of IC3. 

Compared to DP-IC3, which puts all the burden of tightening datapath abstraction 
on refinement's shoulders, applying datapath propagation has two main advantages. 
Firstly, we convey the knowledge of datapath propagations to abstract state space 
by assigning some UFs with accurate values or building the relation between some UFs. 
This knowledge is unknown to the SMT solver but can avoid spurious counterexamples. 
Secondly, we generate DPLs during the datapath propagation. Since datapath propagation 
is independent of CEGAR, combining our method with DP-IC3 can further prune the abstract 
state space and reduce the number of CEGAR iterations. The verification efficiency 
is thus improved.

We develop some strategies to make datapath propagation a lightweight and fast procedure. 
First, the same query formula $\hat{F_k} \wedge \hat{T} \wedge \neg \hat{P}'$ may be called 
many times in IC3 (line 6 in Alg.~\ref{alg:ic3}). Therefore, for each kind of abstract 
query formula $\hat{\varphi}$, if no DPL is generated from the current call of 
Alg.~\ref{alg:dp-propagation}, we skip the datapath propagation on it in later SMT queries. 
Second, we want more concise datapath lemmas. If the size of a query formula is too big, 
the generated DPL may be too long and thus too weak to eliminate spuriousness. Therefore, 
we set a maximal propagation depth $\textit{bound}$ (currently 20) to limit the search 
depth. Actually, in most cases, the datapath propagation terminates within $\textit{bound}$ 
for each abstract query formula $\hat{\varphi}$; either because it finds DPL $\hat{\psi}$ that 
$\hat{\varphi} \wedge \hat{\psi}$ is \emph{unsat}, or no more propagations can be conducted. 

\subsubsection*{\textup{\textbf{Example3}}}
Considering the example design in Fig.~\ref{fig:example}, we show how the datapath 
propagation works with the DP-IC3 algorithm. First, the 0-step abstract query formula 
$\hat{\varphi} = \hat{I}\wedge \neg\hat{P}$, i.e., $\hat{x}=\hat{0} \wedge \hat{y}=\hat{0} 
\wedge \neg \textit{LE}(\hat{y}, \hat{x})$. If $\hat{\varphi}$ is satisfiable, 
$\hat{x} = \hat{0}$ and $\hat{y} = \hat{0}$ should satisfiable too. Therefore, 
we propagate $\hat{0}$ to $\textit{LE}(\hat{y}, \hat{x})$ and obtain $\textit{LE}(\hat{0}, 
\hat{0})$. Since $\models_{\mathcal{T}} 0 \leq 0$, we have $\textit{LE}(\hat{0}, \hat{0})$ 
is $\textit{true}$. Consequently, its negation is propagated to $\textit{false}$. 
Datapath propagation returns $\textit{unsat}$ for $\hat{\varphi}$ and generate DPL 
$\textit{LE}(\hat{0}, \hat{0})$. Additionally, since $\hat{x}$ and $\hat{y}$ are 
in the same equality closure, we obtain another DPL $\hat{x}=\hat{y} \rightarrow 
\textit{LE}(\hat{y}, \hat{x})$. 

The 1-step abstract query formula $\hat{\varphi}=\hat{I}\wedge\hat{T}\wedge\neg\hat{P}$:
\begin{equation*}
    \begin{split}
        &\hat{x}=0\wedge\hat{y}=0 \wedge \hat{x}' = \textit{LT}(\hat{x}, \hat{y})? 
        \hat{x}:(\hat{y}!=\hat{x})?\hat{y}:\textit{ADD}(\hat{x}, \hat{1}) 
        \wedge \\ &\hat{y}' = (\hat{y} == \hat{x})?\textit{ADD}(\hat{y}, \hat{1}):
        \textit{LT}(\hat{x}, \hat{y})?\hat{y}:\hat{x} \wedge 
        \neg \textit{LE}(\hat{y}', \hat{x}')
    \end{split}
\end{equation*}
First, the literals $\hat{x} = \hat{0}$ and $\hat{y} = \hat{0}$ are deduced $true$. 
Then, $\hat{0}$ is propagated to related UFs. Since $\models_{\mathcal{T}}\neg(0<0)$, 
$\textit{LT}(\hat{x}, \hat{y})$ is propagated to $\textit{false}$. Meanwhile, 
$\hat{y} != \hat{x}$ is also $\textit{false}$. $\hat{\varphi}$ is updated to 
$\hat{x}'=\textit{ADD}(\hat{0},\hat{1}) \wedge \hat{y}'=\textit{ADD}(\hat{0},\hat{1})
\wedge \neg \textit{LE}(\hat{y}',\hat{x}')$. Next, since $\models_{\mathcal{T}} 0+1=1$ 
and $\hat{1} \in symb(\hat{\varphi})$, we replace $\textit{ADD}(\hat{x}, \hat{1})$ 
with $\hat{1}$ and we get DPL $\textit{ADD}(\hat{0}, \hat{1}) = \hat{1}$. 
$\hat{\varphi}$ is updated to $\hat{x}'=\hat{1} \wedge \hat{y}'=\hat{1} 
\wedge \neg \textit{LE}(\hat{y}',\hat{x}')$. Since $\models_{\mathcal{T}} 1\leq 1$, 
$\textit{LE}(\hat{y}',\hat{x}')$ is propagated to $\textit{true}$. Meanwhile, 
we get DPLs $\textit{LE}(\hat{1}, \hat{1})$ and $\hat{x}'=\hat{y}' \rightarrow 
\textit{LE}(\hat{y}', \hat{x}')$. Finally, $\neg \textit{LE}(\hat{y}',\hat{x}')$ 
is deduced to $fasle$, datapath propagation finds $\hat{\psi}$ that $\hat{\varphi} 
\wedge \hat{\psi}$ is \emph{unsat} and returns $\textit{unsat}$ for $\hat{\varphi}$. 
Therefore, datapath propagation reports $\textit{unsat}$ to DP-IC3 and combines 
$\hat{\psi}$ to $\Phi_{\textit{dpl}}$. 

The next abstract query formula $\hat{\varphi} = \hat{P} \wedge \hat{T} 
\wedge \neg \hat{P}'$. Since $y \leq x \models_{\mathcal{T}} \neg (x <y)$, 
$\textit{LT}(\hat{x}, \hat{y})$ is propagated to $\textit{false}$. Then, 
no more propagations can be conducted because $\hat{\varphi}_p = \hat{\varphi}_q$ 
after the first iteration in Alg.~\ref{alg:dp-propagation}. Therefore, 
$\hat{\varphi}\wedge \Phi_{\textit{drl}} \wedge \Phi_{\textit{dpl}}$ is 
passed to SMT solver. Finally, an inductive invariant $\hat{y} = \hat{x}$ 
is found after a few rounds of SMT queries. To verify the example design, 
DP-IC3 invokes six refinements in the CEGAR framework, but applying datapath 
propagation generate DPL $\hat{\psi}$ beyond CEGAR and does not call refinement. 
\section{Evaluation}
\label{sec:evaluation}

This section introduces the implementation of our approach and reports the 
comparative results and analysis with some state-of-the-art verification tools. 

\begin{table*}[htb]
    \caption{Summary of experimental results}
    \centering
    \begin{tabular}{lccccccccc}
    \toprule
    \multirow{3}{*}{Verifier} & \multirow{3}{*}{Total} & \multirow{3}{*}{Timeout} & \multicolumn{4}{c}{Verified}  & \multicolumn{2}{c}{Both Verified}   \\ \cmidrule(lr){4-7} \cmidrule(lr){8-9}
                              &                & & \multirow{2}{*}{Num} & \multirow{2}{*}{CPU-Time (s)} & \multirow{2}{*}{Safe} & \multirow{2}{*}{Unsafe}  & \multirow{2}{*}{Num}    & CPU-Time (s)  \\ 
                              &                & &                      &                           &                       &                          &        & (-/\ours)  \\
    \cmidrule(lr){1-9}                                                                                  
    \ours            & \multicolumn{1}{c}{1089} & \multicolumn{1}{c||}{296}     & 793   & 37918.8    & 762  & \multicolumn{1}{c||}{31}      & -         & -                  \\ 
    \avr             & \multicolumn{1}{c}{1089} & \multicolumn{1}{c||}{383}     & 706   & 57483.4    & 675  & \multicolumn{1}{c||}{31}      & 689       & \multicolumn{1}{l}{(38723.1/26538.0, \ 1.46x)}    \\ 
    \ia              & \multicolumn{1}{c}{1089} & \multicolumn{1}{c||}{771}     & 314   & 121810.1   & 286  & \multicolumn{1}{c||}{27}      & 235       & \multicolumn{1}{l}{(98675.7/4829.8, \ 20.40x)}     \\ 
    \pono(ic3sa)     & \multicolumn{1}{c}{1089} & \multicolumn{1}{c||}{420}     & 104   & 11875.7    & 93   & \multicolumn{1}{c||}{11}      & 98        &  \multicolumn{1}{l}{(10895.6/1106.6, \ 9.85x)}     \\ 
    \pono(mbic3)      & \multicolumn{1}{c}{1089} & \multicolumn{1}{c||}{460}     & 66    & 17883.4    & 54   & \multicolumn{1}{c||}{12}      & 58        & \multicolumn{1}{l}{(16830.1/1468.5, \ 11.46x)}     \\ 
    \pono(ind)       & \multicolumn{1}{c}{1089} & \multicolumn{1}{c||}{817}     & 248   & 1867.1     & 239  & \multicolumn{1}{c||}{9}       & 237       & \multicolumn{1}{l}{(1767.6/1568.8, \ 1.13x)}     \\ 
    \pono(sygus-pdr) & \multicolumn{1}{c}{1089} & \multicolumn{1}{c||}{437}     & 84    & 37702.9    & 75   & \multicolumn{1}{c||}{9}      & 84        & \multicolumn{1}{l}{(37702.9/7342.1, \ 5.14x)}     \\
    \bottomrule
\end{tabular}
    \label{tab:results-all}
\end{table*}

\subsection{Implementation and Setup}

We implemented our approach in \avr with around 8K lines 
C++ codes\footnote{Artifact is available: 
\url{https://doi.org/10.5281/zenodo.7333164}}. \avr is 
a state-of-the-art hardware model checker for verifying 
safety properties. We integrate the datapath propagation 
and lemma generation procedures into the original 
verification framework. Our implementation is called \ours. 
We compare \ours with recent well-known hardware verification tools:
\begin{itemize}
    \item \avr\footnote{https://github.com/aman-goel/avr/commit/dbc3371}:
          a tool that implements the IC3-style reachability checking 
          with syntax-guided abstraction (SA) and datapath abstraction; 
          It is the champion tool of the latest hardware model checking 
          competition (HWMCC-2020)\footnote{https://fmv.jku.at/hwmcc20/}. 
    \item \pono (also known as CoSA2)\footnote{https://github.com/upscale-project/pono/commit/b243ce}:
          an SMT-based model checker that implements various reachability 
          checking techniques. It is the champion tool of HWMCC-2019\footnote{https://fmv.jku.at/hwmcc19/}.
    \item \ia \footnote{https://es-static.fbk.eu/people/griggio/ic3ia/index.html}:
          a tool that implements implicit predicate abstraction. 
          It performs reachability checking at the boolean level 
          of the abstract state and eliminates spurious counterexamples 
          by adding a sufficient set of new predicates. 
\end{itemize}
For \avr and \ia, we use their default configurations. 
Since \pono is a platform that implements various techniques, 
we compare \ours with \pono under four different engines: 
1) \texttt{ic3sa} - a basic implementation of IC3 with syntax-guided
abstraction; 2) \texttt{mbic3} - a naive model-based IC3 lifted 
to SMT, which learns clauses of equality between variables 
and model values; 3) \texttt{ind} - k-induction based verification; 
4) \texttt{sygus-pdr} - a implementation of IC3 that employs 
syntax-guided synthesis for lemma generation. 

We collect all the verification tasks of the last two HWMCCs as benchmarks. 
There are 618 and 632 tasks in HWMCC 2019 and HWMCC 2020, respectively. 
After eliminating duplicate cases, we attain 1089 benchmarks. All the 
benchmarks are written in BTOR2\cite{btor2} format, an intermediate 
language for verification, and can be synthesized from Verilog by the 
Yosys\cite{yosys} toolchain. Note that \ia only supports VMT format, 
an extension of SMT-LIBv2; we utilize vmt-tools 
\footnote{http://es-static.fbk.eu/people/griggio/ic3ia/vmt-tools-latest.tar.gz} 
to translate BTOR2 files into VMT files. For \avr and \pono, we use 
1089 BTOR2 files as input. For \ia, we use 1089 VMT files as input. 

All the experiments are conducted on a server with AMD EPYC 7H12 128-core 
CPU and 1TB memory, and the operating system is Ubuntu 20.04 LTS. 
Following the competition, the timeout for each verification task 
is set to 3600 seconds. 

\subsection{Overall Experimental Results}

Tab.~\ref{tab:results-all} summarizes the results of the above tools 
and \ours on all the benchmarks. Columns \texttt{Total} and \texttt{Timeout} 
lists the number of collected benchmarks and cases that exceed the time limit, 
respectively. Columns 4-8 display the data about the verified cases, 
where \texttt{Num} is the number of verified cases, and \texttt{CPU-Time} 
is the accumulated wall clock time, \texttt{Safe} and \texttt{Unsafe} 
are the number of cases that satisfy or violate the specified safety property. 
The column \texttt{Unique} lists the number of cases that can only 
be verified by the selected tool and \ours. The last two columns display 
statistics of tasks that can be verified by the listed tool and \ours. 

\begin{figure}[tb]
    \centering
    \includegraphics[width=\linewidth]{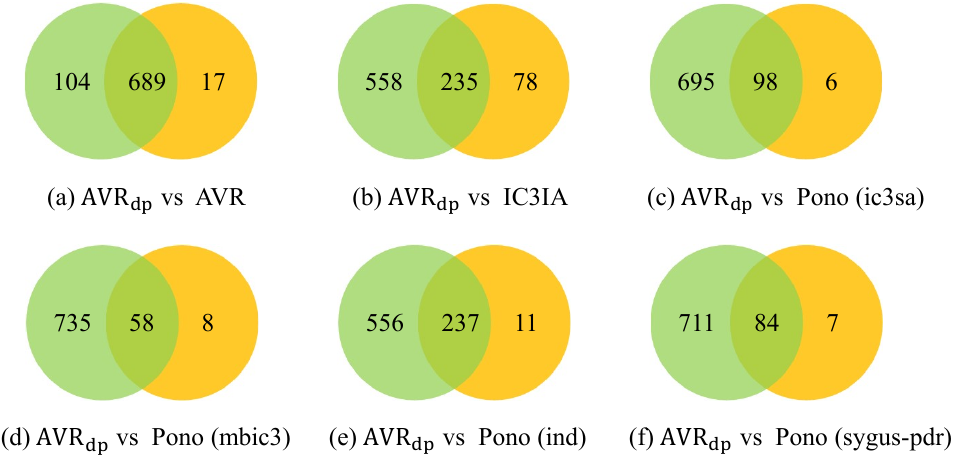}
    \caption{The number of verified cases of \ours (the green circle) 
    and the comparison tool (the yellow circle) where the intersecting 
    region represents cases that both tools can verify.}
    \label{fig:unique}
\end{figure}

There are 1089 benchmarks in total. \avr verifies 706 cases in 57483.4 
seconds, and \ours verifies 793 cases in 37918.9 seconds -- \ours verifies 
87 more cases and achieves 1.52x speedup than \avr. Both \avr and \ours 
can verify 689 cases. Considering these cases, \avr spends 38723.1 
seconds whereas \ours costs 26538.0 seconds -- \ours is 1.46x times 
faster than \avr to verify these same cases. The third row of 
Tab.\ref{tab:results-all} shows the comparative results of \ia 
and \ours. \ia verifies 314 tasks in 121810.1 seconds and timeout 
for 771 tasks. Considering the 235 both-verified cases, \ia spends 
98675.7 whereas \ours only cost 4829.8 seconds 
-- \ours is 20.04x faster than \ia. 

The last four rows display the comparative results of \pono and \ours. 
We use the engine name for brevity. \texttt{ic3sa} verifies 106 cases 
in 11875.7 seconds and considers 98 both-verified cases, \texttt{ic3sa} 
costs 10895.6 seconds and that number of \ours is 1106.6 seconds -- \ours 
is 9.85x faster than \pono with \texttt{ic3sa} engine. \texttt{mbic3} 
verifies 66 cases in 17883.4 seconds, and among 58 both-verified cases, 
the consuming time for \texttt{mbic3} and \ours are 16830.1 and 1468.5 
seconds -- \ours is 11.46x faster than \pono under \texttt{mbic3} strategy. 
\texttt{ind} performs better than the above engines since it verifies 
248 cases using 1867.1 seconds. Among 237 both-verified cases, \ours 
is slightly superior to \texttt{ind}. The last row shows that 
\texttt{sygus-pdr} verifies 84 tasks in 37702.9 seconds. These cases 
can also be verified by \ours with 7342.1 seconds -- \ours is 5.14x 
faster than \pono under \texttt{sygus-pdr} engine. 

\subsection{Results Analysis}

Fig.~\ref{fig:unique} shows the number of verified cases of \ours 
and the comparison tool. According to (a), 104 cases can only 
be verified by \ours. Our approach utilizes datapath propagation 
to prune abstract state space and generates at least one datapath 
lemma for each of them to guide the verification procedure. 
17 cases are just the opposite; \ours is inferior to \avr on 
these cases. The inferiority is because the abstract query 
formulas are too big in these cases, and no useful datapath lemma 
is generated in the propagation procedure. 

Compared to \ia, 558 cases can only be verified by \ours. \ia times 
out on 553 cases and throws exceptions on the other 5 cases. Note that 
\ia is also time-consuming for the 314 verified cases. This is because 
\ia employs predicate abstraction, which performs IC3 on the boolean 
level of abstract state space. It needs to learn a sufficient set of 
predicates to tighten the abstraction. However, generating predicates, 
especially useful predicates, is not easy. It may require numerous CEGAR 
iterations, and maintaining so many predicates is also a heavy burden 
for the verification procedure. Therefore, \ia is often trapped in 
situations where many arithmetic or bitwise operations are involved. 

\begin{figure}[tb]
    \centering
    \includegraphics[width=.8\linewidth]{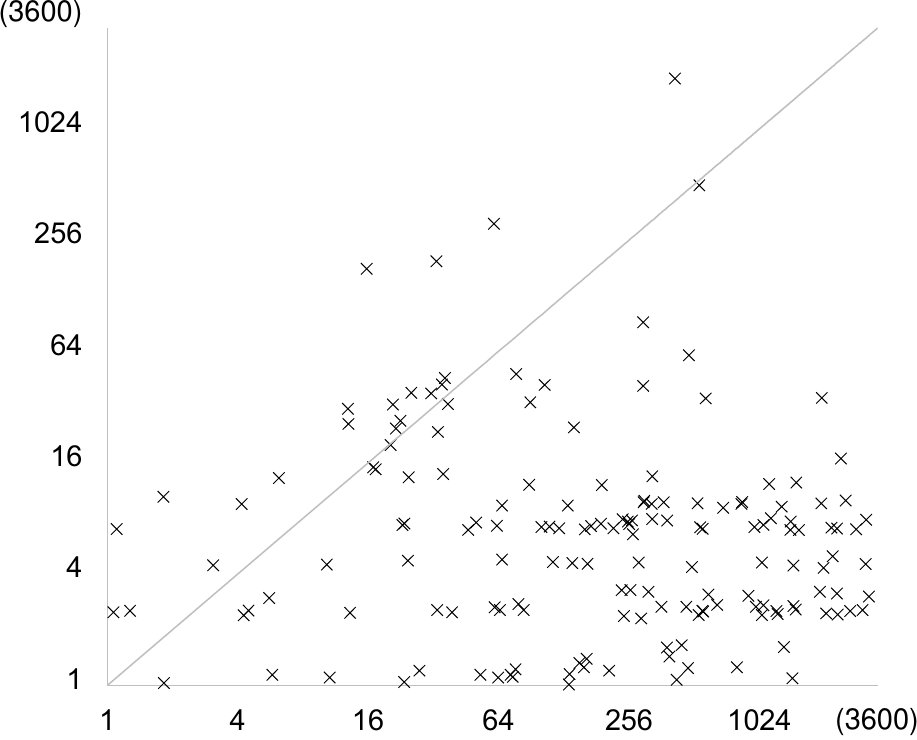}
    \caption{Comparison of \ia ($X$-axis) and \ours ($Y$-axis) in terms of 
    \texttt{CPU-Time} for each task on the benchmarks in Tab.~\ref{tab:result-sel}}
    \label{fig:fig-ia}
\end{figure}

On the contrary, \ours is inferior to \ia on 78 cases since some crucial 
predicates that witness the violation of safety property are found by \ia. 
Fig.~\ref{fig:fig-ia} shows the comparison results of \ia and \ours on 235 
both-verified cases. Each point in the panel corresponds to a verification 
task, with the $X$ and $Y$ coordinates representing the \texttt{CPU-Time} 
of \ia and \ours, respectively. Note that both $x$- and $y$-axis take 
logarithmic coordinates, and each point below/above the diagonal line 
represents a superior/inferior case of our approach against the \ia. 
When the cases become complex, our method starts to show its strength. 
In most cases, our approach is more efficient than \ia. 

Fig.~\ref{fig:unique} (c)-(f) display the number of verified cases of 
\ours and \pono under different engines. 695, 735, 556, and 711 cases 
can only be verified by \ours compared to \texttt{ic3sa}, \texttt{mbic3}, 
\texttt{ind}, and \texttt{sygus-pdr}, respectively. Meanwhile, there are 
around 10 cases that \ours is inferior to the comparison engine of \pono. 
Note that apart from the timeout and verified cases, there are hundreds 
of cases that \texttt{ic3sa}, \texttt{mbic3}, and \texttt{sygus-pdr} 
cannot handle. This is because these engines do not support arrays 
and throw exceptions when they meet arrays. Therefore, we mainly 
focus on the both-verified cases. 

\begin{figure}[tb]
    \centering
    \includegraphics[width=.77\linewidth]{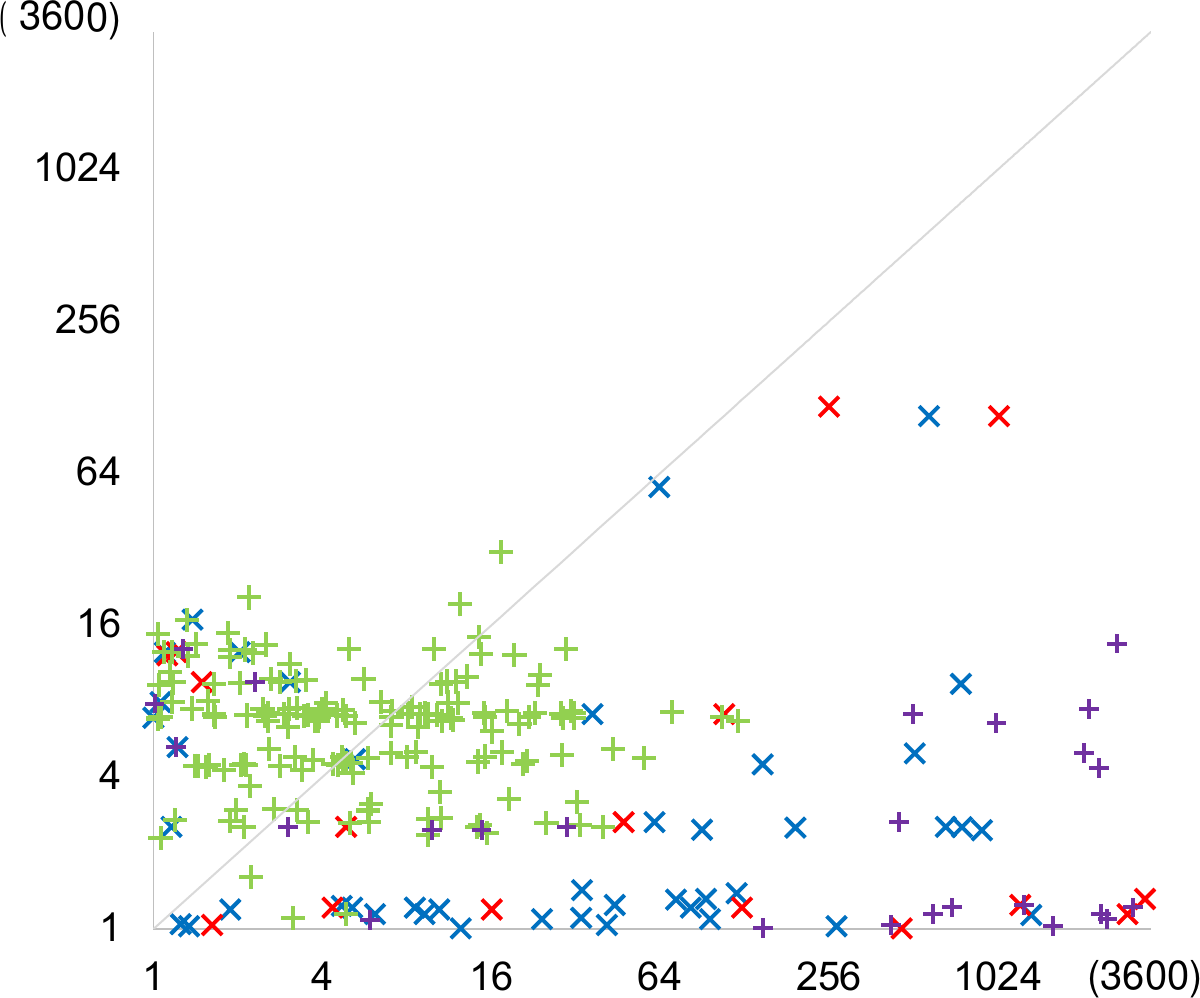}
    \caption{Comparison of \pono ($X$-axis) and \ours ($Y$-axis) in terms of 
    \texttt{CPU-Time} for each task on the full benchmark set. The blue, red, 
    green, and purple points represent the comparison result of \texttt{ic3sa}, 
    \texttt{mbic3}, \texttt{ind}, and \texttt{sygus-pdr}, respectively}
    \label{fig:fig-pono}
\end{figure}

Fig.~\ref{fig:fig-pono} displays the comparison results of \ours 
and \pono on both-verified cases. It is intuitive that \ours is more 
efficient than \texttt{ic3sa}, \texttt{mbic3}, and \texttt{sygus-pdr} 
in most cases because the corresponding points are below the diagonal. 
Compared to \texttt{ind} (green points), the performance of \texttt{ind} 
and \ours is similar. However, these 237 both-verified cases are relatively 
simple since they can be verified by \texttt{ind} and \ours within 
100 seconds. As the complexity and scale of the task increase, \ours 
verifies 556 more cases than \texttt{ind}, which timed out on these cases. 

Note that \avr is the latest champion tool in HWMCC, i.e., it is 
already superior to \pono and \ia. Moreover, we implement the datapath 
propagation and DPL generation in \avr. Therefore, we take \avr as 
a baseline and compare further to show that our approach 
is effective and efficient. \avr eliminates spurious counterexamples 
and tightens the datapath abstraction only by refinement. Instead, 
our approach can generate DPL during the propagation procedure. 
Considering the 689 both-verified cases, \ours generates 3923 
datapath propagation lemmas. These lemmas consider the 
original semantics of datapath operations and tighten the datapath 
abstraction by adding constraints over UFs in the abstract state space. 
Compared to \avr, \ours reduces the number of refinements from 15927 
to 11220 -- \ours has only 70.5\% of the number of refinements of \avr. 

\begin{figure}[tb]
    \centering
    \includegraphics[width=.8\linewidth]{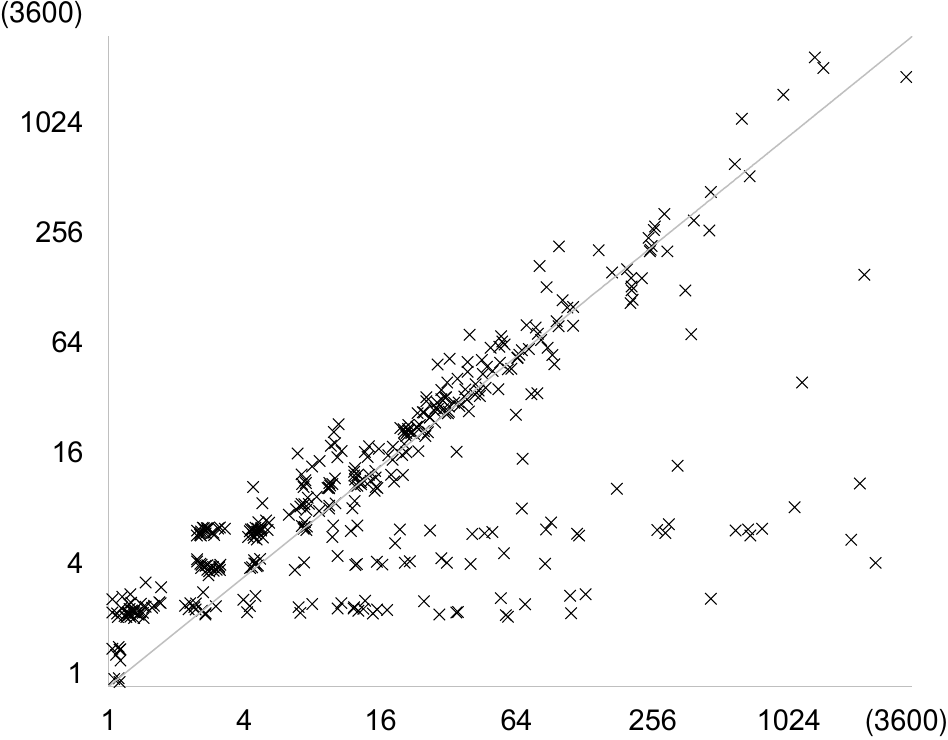}
    \caption{Comparison of \avr ($X$-axis) and \ours ($Y$-axis) in terms of 
    \texttt{CPU-Time} for each task on the full benchmark set}
    \label{fig:fig-avr}
\end{figure}

Fig.~\ref{fig:fig-avr} displays the comparative results of \avr and 
\ours on 689 both-verified cases. The points below the diagonal 
represent the cases that \ours achieves higher efficiency than \avr. 
Among these both-verified cases, there are 372 cases on which 
\ours generates at least one DPL in the propagation for each case 
and 3923 DPLs in total. Tab.~\ref{tab:result-sel} reports the 
statistics of these 372 cases, \avr spends 25186.2 seconds to 
verify these cases, and that number of \ours is 8796.9 seconds 
-- \ours achieves 2.86x speedup than \avr. Moreover, \avr has 
15651 refinements whereas \ours has 10944 -- applying our 
approach reduces 30.1\% refinements on these cases. 

There are 317 both-verified cases that \ours does not generate a
datapath propagation lemma. However, these 317 cases only involve 
276 refinements. After analysis, we found that the initial abstraction 
is accurate enough for these cases to prove correctness or find 
violations. So only a few refinements are called, and there is 
no room for \ours to generate DPL during the propagation. 
These cases correspond to the points in Fig.~\ref{fig:fig-avr} 
on the diagonal or slightly above the diagonal. 

There is a cluster of points in the bottom left of Fig.~\ref{fig:fig-avr} 
These tasks are mostly verified within a few dozen seconds. 
\ours performs similarly to \avr, and some are even inferior. 
Verification time on these tasks is short because they are trivial, 
so results by applying our method are dominated by the time to traverse 
literals in the abstract query formula. However, as the benchmark's 
scale increases, our approach can bring promising speedups. 

\subsection{Discussions}

\subsubsection*{\textbf{Scalability}}
The datapath propagation is orthogonal to the CEGAR framework and attempts 
to generate datapath propagation lemmas over datapath operations. Although 
we focus on the IC3 algorithm within the datapath abstraction and refinement 
framework, the DP-IC3 algorithm serves as an encoder that calls 
abstract formula queries on demand. Datapath abstraction has been widely adopted 
in many fields, such as equivalence checking \cite{eq-check-fmcad16,eq-check-spin13} 
that hides the details of each hardware design and eliminates spurious behaviors 
in refinement iteration. The idea of this paper can be generalized to these problems. 
Instead of taking all the datapath operations as uninterpreted and putting 
the heavy burden on CEGAR, lightweight strategies, or heuristics may convey 
essential information from a different perspective, guide the abstraction-based 
verification, and improve its scalability and efficiency.

\subsubsection*{\textbf{Threats to Validity}}
The main threats to our method's validity are whether the performance 
improvements are due to our tactic and whether our implementation and 
experimental results are credible. Firstly, we implement the proposed 
method in \avr and make a comparison with it. The improvements over 
\avr must come from our tactic. Secondly, the reduction in the number 
of refinements is consistent with the theoretical analysis, which 
confirms that the improvements are indeed from our approach. Thirdly, 
our implementation is loosely coupled with the original verification 
framework. Benchmarks are collected from the latest two HWMCCs, one 
of the most representative and convincing open sources in hardware 
verification. Moreover, we compare our method with the newest version 
of the state-of-the-art tools. We are thus confident in the effectiveness 
of our tactic. 

\begin{table}[t]
    \centering
    \caption{Results for cases that pre-deduction learned at least one lemma}
    \begin{tabular}{cccccc}
    \toprule
    \multirow{2}{*}{Verifier} & \multicolumn{5}{c}{Both-Verified}    \\ \cmidrule{2-6} 
                          & \multicolumn{1}{c}{Num} & \multicolumn{1}{c}{Safe} & \multicolumn{1}{c}{Unsafe} & \multicolumn{1}{c}{CPU-Time} & \multicolumn{1}{c}{Refinement} \\ \cmidrule(lr){1-6}
    \multicolumn{1}{l}{\avr}  & \multicolumn{1}{c}{372} & \multicolumn{1}{c}{366}  & \multicolumn{1}{c}{6}      & \multicolumn{1}{c}{25186.2} & \multicolumn{1}{c}{15651}  \\
    \multicolumn{1}{l}{\ours} & \multicolumn{1}{c}{372} & \multicolumn{1}{c}{366}  & \multicolumn{1}{c}{6}      & \multicolumn{1}{c}{8796.9}  & \multicolumn{1}{c}{10944}  \\
    \bottomrule
    \end{tabular}
    \label{tab:result-sel}
\end{table}

\subsubsection*{\textbf{Limitations}}
The main limitations of our approach are summarized below. 
Firstly, we currently focus on datapath operations about arithmetic, 
relational, bit-wise, shifting, and logical operations. We also develop 
some strategies to make it a lightweight and fast procedure. Therefore, 
our approach is sound but incomplete; it may not generate datapath 
propagation lemmas in some situations. To improve the scalability 
of the proposed method, we plan to elaborate on more propagations 
about arrays, concatenation, and extraction operations. Secondly, 
generalization is a crucial factor in hardware verification. 
Elaborated strategies and heuristics for generalizing datapath 
propagation lemmas are required to improve the overall efficiency further. 
\section{Related Works}
\label{sec:related}

Numerous studies have been widely studied on improving the availability 
and efficiency of the IC3-based approach and applying constant propagation 
to verification. We discuss representative techniques in these two fields. 

\subsection{Advanced IC3-based approach}

IC3 has been the most successful and talented technique for hardware verification 
in recent years. Various optimizations are developed to improve the bit-level 
IC3 engine. PDR\cite{pdr-11,pdr-12} proposes a simplified and faster implementation 
of IC3 by using three-valued simulation. It ignores redundant bit-level details 
to reduce the heavy burden on the SAT solver. In this way, PDR learns short 
clauses without numerous generalizations and achieves a significant speedup. 
UFAR~\cite{UFAR17} is a hybrid word- and bit-level solver that replaces heavy 
bit-level arithmetic logic with UF in Bounded model checking (BMC) ~\cite{bmc00} 
or in PDR. These UFs referring to arithmetic operations are bit-blasted 
and given to a sound and complete bit-level model checker. This hybrid 
approach temporarily neglects the complicated bit-level detail of 
arithmetic operations and demonstrates its scalability. 

The recent work~\cite{cav23-ic3} proposes to search for the so-called 
$i$-good lemmas in bit-level IC3. These lemmas are crucial for refining 
the over-approximating sequence and reaching a fixed point in the safe case. 
Two heuristics are developed to find such lemmas. The \emph{branching} 
heuristic controls how the SAT solver extracts an unsatisfiable core 
by privileging variables in $i$-good lemma. The \emph{refer-skipping} 
heuristic controls lemma generation by avoiding dropping literals occurring 
in a subsuming lemma in the previous lemma. The $i$-good lemmas and proposed 
heuristics result in significant improvements in terms of performance. 
GSpacerBV~\cite{Gspacer20} replaces bit-blasting in PDR with a technique 
for iterative approximate quantifier elimination in BV. The implementation 
for solving constrained horn clauses (CHC) shows competitive performance 
compared to other advanced CHC solvers. 

However, the bit-level IC3 and its variants still suffer from the state 
space explosion problem. Many advanced abstraction techniques are proposed 
to lift the IC3 from bit-level to word-level. IC3IA~\cite{ic3ia-14,ic3ia-16} 
proposed a tight integration of IC3 with implicit abstraction~\cite{IA09}, 
a form of predicate abstraction~\cite{PA01,PA97}. With this technique, 
IC3 operates at the Boolean level of the abstract state space and generates 
inductive clauses over the abstraction predicates. When a spurious 
counterexample permitted by the current abstraction is found, it is 
refined by incrementally generating and adding a set of new predicates. 
However, it takes work to attain useful predicates. IC3IA may learn 
numerous predicates during the search to eliminate spurious counterexamples. 

Averroes~\cite{avr-14} first integrates the IC3 with datapath abstraction. 
The approach can be seen as two layers of the CEGAR loop. The inner loop 
conducts IC3 on the datapath-abstracted state space. The outer loop tightens 
the current abstraction by generating datapath refinement lemmas. These 
datapath lemmas refute the spurious counterexample that the inner loop returns. 
Since datapath operations are essential components in Verilog RTL design, 
the integration helps the verification procedure focus on the big picture 
of the checked property. It cares about the bit-level detail only when 
checking the feasibility of abstract counterexample. However, roughly 
abstracting all the datapath operations as UFs makes the verification 
framework lose all the semantics of datapath operations, which may be 
useful for pruning the abstract state space. Chen~\cite{chen18,tosem18} 
applies the knowledge of the control-flow graph in SMT solving, 
and Zpre~\cite{ppopp22} utilizes the knowledge of thread-interleaving 
to accelerate the concurrent program verification. Both techniques achieve 
promising improvements in terms of efficiency. Inspired by these works 
and considering the knowledge of datapath operations, we propose datapath 
propagation to convey important information to the verification procedure 
and guide the datapath abstraction and refinement. 

AVR\cite{avr-14, avr-tacas20} extends Averroes with syntax-guided 
abstraction (SA). This extension encodes the abstract state space 
using the partition assignment of the set of UFs in the 
word-level syntax. Therefore, IC3 with SA+UF allows for efficient 
reasoning regardless of the bit-width of variables or the complexity 
of datapath operations. AVR is the champion tool in the last two 
HWMCC. SyGuS-APDR~\cite{synatx-guided-lemma21,pono21} utilizes 
syntax-guided synthesis to generate word-level lemmas heuristically. 
It includes a pre-defined grammar template and term production rules 
for generating candidate lemmas. These validated lemmas may prune 
the bad state space and tighten previous frontiers. 

\subsection{Constant propagation in Verification}

Constant propagation is an optimization technique commonly 
used in compilers and software analysis tools. The main goal 
is to replace variables or expressions with their constant 
values wherever possible, effectively reducing the complexity 
of the system representation. Regarding formal verification, 
constant propagation can simplify the verification process 
and reduce the state space, leading to more efficient verification. 

Armoni.et.al~\cite{deeperBMC07} use constant propagation to iteratively 
simplify the formulas submitted to the SAT solver. They build a directed 
acyclic graph, called expression graph (EG), for each BMC formula. 
The propagation starts from the leaves that denote variables or constants 
and updates the EG dynamically. Since BMC instants involve variables 
in different time frames, constant propagation is useful for pruning 
numerous variables and reduces the complexity of the logical expressions. 
Wegman.et.al~\cite{wegman1991constant} propose elaborated algorithms 
in flow analysis. With this technique, constants within conditional 
statements can be propagated if the conditions guarantee that a certain 
variable will always have a constant value under those conditions. 
This can lead to the elimination of branches and further state space reduction. 

Different from the above applications, we focus on datapath operations 
in Verilog RTL design and perform constant propagation across concrete 
and abstract state space. In this way, we consider the original semantics 
of datapath operations, attain their outcomes, and propagate these results 
to corresponding UFs iteratively in abstract state space. 
\section{Conclusion}
\label{sec:conclusion}

In this paper, we presented a datapath propagation mechanism for datapath 
abstraction-based hardware verification. We leverage concrete constant values 
to iteratively compute the outcomes of relevant datapath operations and their 
associated uninterpreted functions in the abstract state space. Meanwhile, we 
generate datapath propagation lemmas in abstract state space and tighten the 
datapath abstraction. 

We implemented the proposed method in a prototype tool named \ours and conducted 
experiments to compare \ours with state-of-the-art hardware verification tools. 
We collected 1089 benchmarks from hardware model checking competition 2019-2020. 
The experimental results show that our approach is effective and efficiency.

\bibliographystyle{IEEEtran}
\bibliography{citations}

\end{document}